\documentclass[sigconf, 10pt]{acmart}
\usepackage{subcaption}
\usepackage{multirow}
\usepackage{booktabs}
\usepackage{array}
\usepackage{graphicx}   
\usepackage{makecell}   
\usepackage{xcolor}     
\usepackage{pifont}    
\usepackage{tabularx}
\usepackage{tcolorbox}

\newcommand{\cmark}{\textcolor{green!60!black}{\ding{51}}}%
\newcommand{\xmark}{\textcolor{red!60!black}{\ding{55}}}%

\AtBeginDocument{%
  }
\setcopyright{none}
\copyrightyear{2018}
\acmYear{2018}
\acmDOI{XXXXXXX.XXXXXXX}
\acmConference[]{}{}{}
\acmISBN{}
\begin{document}
\title{SynthRM: A Synthetic Data Platform for Vision-Aided Mobile System Simulation}
\author{Yingzhe Mao}
\affiliation{%
  \institution{Sun Yat-sen University}
  \city{Guangzhou}
  \country{China}}
\email{maoyzh7@alumni.sysu.edu.cn}

\author{Chao Zou}
\affiliation{%
  \institution{Sun Yat-sen University}
  \city{Guangzhou}
  \country{China}}
\email{zouch28@mail2.sysu.edu.cn}

\author{Yanqun Tang}
\affiliation{%
  \institution{Sun Yat-sen University}
  \city{Guangzhou}
  \country{China}}
\email{yqtang8@mail.sysu.edu.cn}

\renewcommand{\shortauthors}{Yingzhe et al.}
\def\grant{This work was in part by the Science and Technology Planning Project of Key Laboratory of Advanced IntelliSense Technology, Guangdong Science and Technology Department under Grant 2023B1212060024.}

\newcolumntype{C}{>{\centering\arraybackslash}X}
\newcolumntype{P}{>{\centering\arraybackslash}p{0.8cm}}

\newcommand{\nosection}[1]{\vspace{3pt}\noindent\textbf{#1.}}
\newcommand{\nosubsection}[1]{\vspace{3pt}\noindent$\bullet$\hspace{1mm}\textbf{#1}}

\definecolor{myblue}{RGB}{204, 238, 238}
\newtcolorbox{qbox}{colback=myblue!20!white, colframe=myblue!60!black, boxrule=0.2pt, arc=4pt, left=1em, right=1em, top=0.2em, bottom=0.2em, after skip=1em, before skip=0.5em}

\tolerance=1000
\setlength{\parskip}{0pt}
\begin{abstract}
Vision-aided wireless sensing is emerging as a cornerstone of 6G mobile computing. While data-driven approaches have advanced rapidly, establishing a precise geometric correspondence between ego-centric visual data and radio propagation remains a challenge. Existing paradigms typically either associate 2D topology maps and auxiliary information with radio maps, or provide 3D perspective views limited by sparse radio data. This spatial representation flattens the complex vertical interactions such as occlusion and diffraction that govern signal behavior in urban environments, rendering the task of cross-view signal inference mathematically ill-posed. To resolve this geometric ambiguity, we introduce SynthRM, a scalable synthetic data platform. SynthRM implements a Visible-Aligned-Surface simulation strategy: rather than probing a global volumetric grid, it performs ray-tracing directly onto the geometry exposed to the sensor. This approach ensures pixel-level consistency between visual semantics and electromagnetic response, transforming the learning objective into a physically well-posed problem. We demonstrate the platform's capabilities by presenting a diverse, city-scale dataset derived from procedurally generated environments. By combining efficient procedural synthesis with high-fidelity electromagnetic modeling, SynthRM provides a transparent, accessible foundation for developing next-generation mobile systems for environment-aware sensing and communication.
\end{abstract}


\begin{teaserfigure}
  \includegraphics[width=\textwidth]{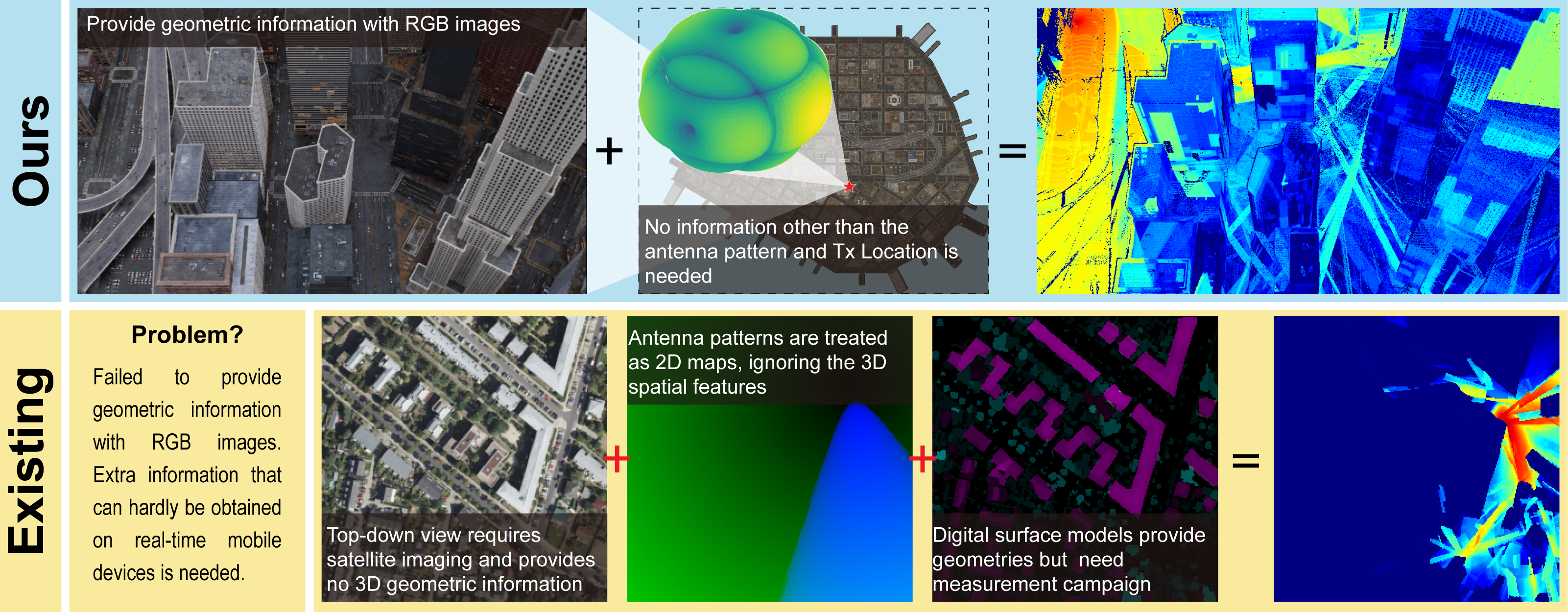}
  \caption{A glance of what SynthRM offers in terms of visual-radio alignment, and what it solves.}
  \Description{A glance of what SynthRM offers in terms of visual-radio alignment, and what it solves.}
  \label{fig:teaser}
\end{teaserfigure}
\received{5 December 2025}
\settopmatter{printfolios=true}
\settopmatter{printacmref=false}
\renewcommand\footnotetextcopyrightpermission[1]{}
\pagestyle{plain}
\maketitle

\section{Introduction}

The transition to 6G is driving a fundamental architectural shift in mobile computing: the dissolution of the boundary between sensing and communication \cite{lin20236g}. As wireless networks migrate to millimeter-wave (mmWave) and terahertz (THz) frequencies, the channel becomes quasi-optical, exhibiting extreme sensitivity to blockage and requiring precise directional alignment \cite{heath2016overview, rangan2014millimeter, ullah2023survey}. In this high-mobility regime, the separation between \textit{seeing} the environment and \textit{communicating} through it is dissolving. Consequently, the field is increasingly adopting Vision-Aided Wireless Sensing, where visual semantics (RGB, Depth) serve as a robust proxy for Channel State Information (CSI), enabling applications ranging from proactive beam management to immersive digital twins.

To fully realize the potential of vision-aided 6G, the underlying data infrastructure must accurately reflect the complex interplay between scene geometry and signal propagation. Predominant methodologies often utilize 2D \textit{Allocentric} (top-down) heatmaps to represent radio coverage. While effective for macroscopic network planning, this representation is essentially a lossy compression of the 3D electromagnetic field, flattening complex vertical features such as building facades and street canyons that are physically critical for ground-based agents. When paired with \textit{Ego-centric} (perspective) imagery, this creates a fundamental geometric schism: the 2D target map lacks the vertical granularity to represent façade-level interactions, while the visual input is naturally restricted by occlusion. Attempting to regress a global, completion-based radio map from a single, partial visual observation forces the model to resolve ambiguities without sufficient causal data, rendering the cross-modal inference task mathematically ill-posed.

We propose that to train robust neural estimators, the learning problem must be reformulated to be physically well-posed. This can be achieved by aligning the simulation domain with the sensor's perception manifold. By restricting the radio simulation to the exact surfaces visible to the camera, we establish a direct, pixel-level causality between the optical features (texture, material, depth) and the electromagnetic response (reflection, scattering). This ``Visible-Aligned-Surface'' (VAS) strategy transforms the learning objective: instead of memorizing site-specific global maps, the model is incentivized to learn generalizable physical laws such as the attenuation coefficients of specific materials or the diffraction patterns at visible edges, which can subsequently be applied to infer environmental characteristics in novel scenarios.

To support this direction, we introduce SynthRM, a scalable, open-source synthetic data platform\footnote{Project can be found at https://github.com/Myzz2003/SynthRM-Platform}. SynthRM utilizes a procedural pipeline to generate diverse urban topologies and implements the VAS simulation strategy to establish high-fidelity geometric correspondence. By leveraging depth-based back-projection, our system reconstructs the geometric manifold exposed to the sensor's view frustum (the VAS) and performs ray-tracing specifically on these visible surfaces using the Sionna-RT backend \cite{sionnart2023, Jakob2020DrJit}. This process generates a \textit{radio-textured} surface, ensuring that the resulting data serves as a spatially accurate representation of signal propagation relative to the visible scene geometry, capturing the precise effects of distance, incidence angle, and structural occlusion.

SynthRM is designed to provide a platform for next-generation channel modeling. To facilitate broad access to high-fidelity simulation, we integrate an optimized, open-source toolchain capable of generating city-scale environments on consumer-grade hardware. Furthermore, to support emerging 3D structure learning frameworks such as Neural Radiance Fields \cite{nerf22023,wrf2025} and 3D Gaussian Splatting \cite{radsplatter2025,gsrf2025}, the platform natively exports precise 3D structural metadata, including camera intrinsics, extrinsics, and reconstructed scene polygons, alongside RGB imagery.

In summary, our contributions are as follows:
\begin{itemize}
    \item \textbf{A Visible-Surface-Aligned Architecture:} We introduce a simulation pipeline that maps radio propagation directly onto camera-visible meshes. This provides a strong geometric prior that resolves the ill-posed nature of cross-view signal inference, enabling rigorous pixel-to-signal learning.
    \item \textbf{Accessible Platform:} We leverage procedural generation and an open-source ray-tracing backend to create infinite, diverse urban topologies, enabling the community to generate massive multi-modal datasets free from commercial licensing barriers and computable on standard consumer-grade electronics.
    \item \textbf{3D-Native Data Structure:} We provide the first radio map dataset to include explicit 3D geometric metadata and uncompressed imagery, enabling the development of geometrically consistent, environment-aware communication models.
\end{itemize}
\section{Motivation}
\label{sec:motivation}
The development of robust neural estimators for radio environments requires a formulation that respects the underlying physics of electromagnetic propagation \cite{kim2025efficient}. In this section, we analyze the causal link between site geometry and signal behavior. We first define the geometric primitives that govern propagation mechanisms, then demonstrate how existing \textit{mismatched} dataset formulations sever this link, creating an ill-posed inverse problem that necessitates the spatially aligned approach proposed in SynthRM.

\begin{table*}[htbp]
\centering
\caption{Geometric Determinants of Radio Propagation: Comparison of Information Availability in Datasets}
\label{tab:geo_factors}
\renewcommand{\arraystretch}{1.2}
\begin{tabular}{p{0.18\linewidth} c p{0.48\linewidth} c c}
\toprule
\textbf{Geometrical Factor} & \textbf{Symbol} & \textbf{Propagation Role \& Physical Governance} & \textbf{Mismatched} & \textbf{Aligned} \\
\midrule
Path Length & $d$ & Dictates FSPL ($\propto 1/d^2$), the baseline energy budget. & Partial & Full \\
Surface Normal & $\mathbf{n}$ & Defines incidence plane for Snell’s Law and Fresnel Equations. & Partial & Full \\
Angle of Incidence & $\theta_i$ & Inputs to Fresnel coefficients for magnitude and phase shift. & Partial & Full \\
Interaction Count & $N_{\text{int}}$ & Determines cumulative path loss and total phase delay. & Partial & Full \\
Obstacle Geometry & $\mathcal{G}_{xyz}$ & Defines ray path validity (LOS vs. blocking) and interaction points. & Partial & Full \\
Edge Geometry & $\alpha$ & Inputs to UTD for diffraction into shadowed (NLOS) regions. & Partial & Full \\
\bottomrule
\end{tabular}
\\ \vspace{0.5ex}
\raggedright\footnotesize{\textit{Note: "Mismatched" refers to datasets pairing ego-centric views with global/allocentric maps. "Aligned" refers to spatially consistent view-map pairs (Proposed). "Partial" indicates the information is frequently occluded or inferred without causal data.}}
\end{table*}

\subsection{The Geometric Determinism of Radio Propagation}
Radio wave propagation is governed by the deterministic interaction between electromagnetic waves and the 3D environment. As established in ray optics, the received signal is a superposition of multipath components, each generated by specific geometric primitives---surfaces, edges, and wedges. Consequently, the problem of estimating a radio map from visual data is effectively an inverse rendering problem: to correctly infer the signal field, the model must implicitly invert the physical interactions defined in Table~\ref{tab:geo_factors}.

\subsection{The Ill-Posed Nature of Mismatched Alignment}
In critically analyzing existing vision-aided radio map datasets, we identify a fundamental misalignment between the input modality and the prediction target. The prevailing paradigm pairs ego-centric visual inputs (e.g., sparse views from a UAV, car, or robot) with global, allocentric radio maps (top-down coverage of the entire area). As summarized in the \textit{Mismatched} column of Table~\ref{tab:geo_factors}, this formulation renders the reconstruction problem mathematically ill-posed.

The core issue is the unavailability of latent geometric states. In an ego-centric view, the sensor is subject to perspective projection and occlusion; it captures only the visible surfaces (Partial $\mathcal{G}_{xyz}$ and $\mathbf{n}$), whereas the target radio map covers occluded regions driven by \textit{invisible} geometry. Specifically, radio waves propagate into shadowed regions via diffraction at structural edges (Edge Geometry $\alpha$). In a mismatched setup, a camera viewing a building facade cannot observe the diffracting corners required to predict the field behind that structure. Similarly, the total received energy is a function of the accumulated path length ($d$) and the interaction history ($N_{\text{int}}$); when reflection paths rely on surfaces outside the camera's frustum, the model lacks the causal inputs necessary to compute path loss or phase delay.

Under these conditions, neural networks cannot learn the physical priors of propagation. Instead, they rely on statistical hallucination---guessing the signal distribution based on dataset biases (e.g., ``buildings usually cast shadows'') rather than causal reasoning. This dependency limits generalization, as the model will fail in environments where the hidden geometry deviates from the training distribution.

\subsection{Proposal: The Simulation on VAS Paradigm}
To resolve this ambiguity, we propose the \textbf{Simulation on VAS} paradigm, realized through the SynthRM platform. Unlike previous mismatched approaches, the VAS strategy aligns the visual frustum with the radio simulation boundaries. By ensuring that every pixel in the predicted radio map corresponds to a spatial volume causally linked to the visible geometry, we convert the ill-posed problem of hallucination into a well-posed problem of physical inference.
\section{System Design}
\label{sec:system_design}

To address the ill-posed nature of radio map reconstruction identified in Section \ref{sec:motivation}, we introduce \textbf{SynthRM}, a modular simulation platform designed to enforce strict geometric consistency between visual perception and radio propagation. Unlike traditional pipelines that simulate radio coverage on a global grid and loosely pair it with visual data, SynthRM adopts a surface-centric approach. It explicitly reconstructs the geometry visible to the sensor and performs ray-tracing directly on the manifold of that visible geometry.

\subsection{Architectural Overview}
The core philosophy of SynthRM is the ``Simulation on VAS'' paradigm. Standard radio mapping typically employs volumetric probing, where signal strength is estimated at regular intervals across a global coordinate system. While computationally straightforward, this approach decouples the signal from the physical surfaces that govern its propagation, leading to the geometric disconnect discussed earlier.

\begin{figure*}[htbp]
    \centering
    \includegraphics[width=\textwidth]{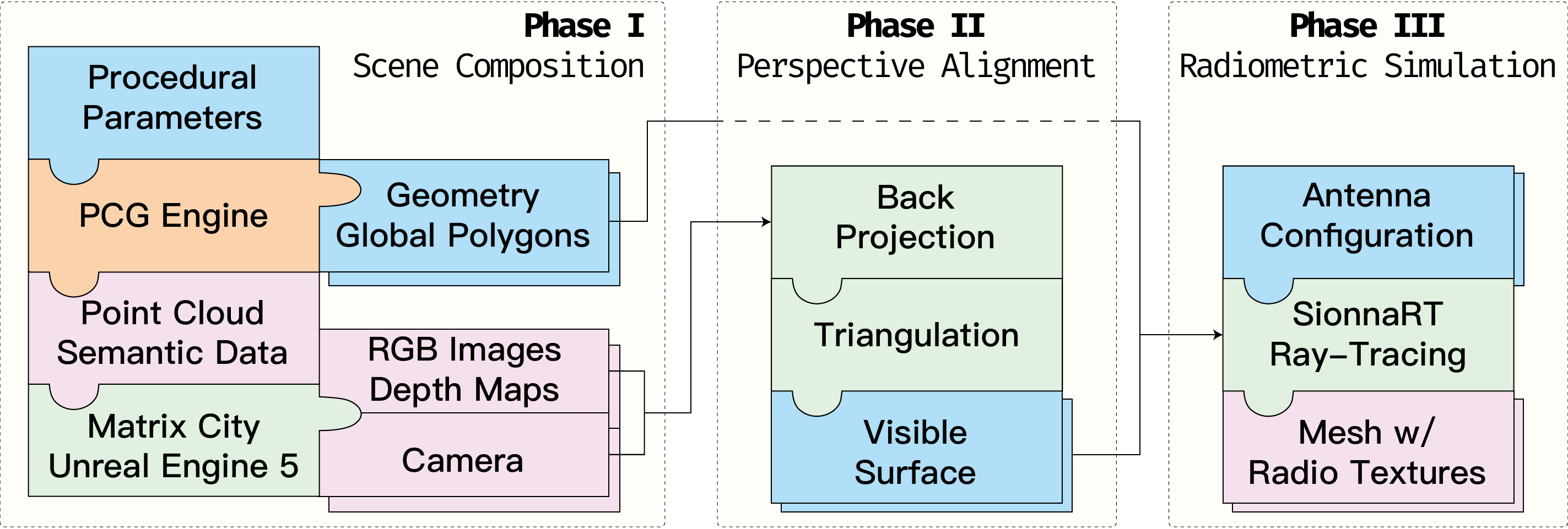}
    \caption{\textbf{The SynthRM Data Collection Workflow.} The pipeline transitions from procedural scene generation (Composition) to frustum-specific mesh recovery (Alignment), culminating in surface-bound ray tracing (Simulation). This ensures radio metrics are topologically mapped to the visible surfaces.}
    \label{fig:workflow}
\end{figure*}

In contrast, our architecture binds the simulation domain to the sensor's view frustum. By defining the simulation targets as the surfaces visible to the camera, we ensure that the resulting radio map is intrinsically aligned with the visual input. As illustrated in Figure~\ref{fig:workflow}, the pipeline executes as a linear sequence of three transformations:
\begin{enumerate}
    \item \textbf{Scene Composition:} Urban environments are synthesized via Procedural Content Generation (PCG) to produce diverse, geometrically complex topologies.
    \item \textbf{VAS Reconstruction:} The raw depth buffers and camera extrinsics are used to back-project the visual data into a mesh that strictly represents the camera's field of view.
    \item \textbf{Mesh-Centric Radio Simulation:} Finally, the system executes ray-tracing on the reconstructed manifold. Instead of probing a volumetric grid, receivers are instantiated at the face centers of the visible mesh, binding electromagnetic properties directly to the surface topology.
\end{enumerate}
This progression guarantees that the radio signal is treated as a material property of the observed geometry rather than a detached spatial field.

\subsection{Scene Composition}
SynthRM harnesses PCG to synthesize a diverse array of urban environments. We utilize a node-based procedural modeling architecture that enables the construction of intricate urban topologies—including road networks, zoning districts, and building footprints—with minimal manual intervention. This parametric approach allows for the generation of heterogeneous urban scenarios by simply tuning high-level attributes such as building density and street width.

The procedural output is bifurcated into two distinct data streams to serve both modalities. The first is the raw 3D geometry (polygons) used for physical interactions. The second consists of attribute-rich 3D point clouds designed for optical rendering. To bridge the gap between geometric abstraction and photorealism, we employ the high-fidelity Unreal Engine 5 together with the Matrix City plugin \cite{li2023matrixcity} to dynamically instantiate cinematic-quality textures and assets upon the coarse underlying skeleton. This dual-stream pipeline allows for the synthesis of movie-level imagery without the prohibitive memory overhead associated with storing explicit high-poly city models.

\begin{figure}[htbp]
    \centering
    \includegraphics[width=0.45\textwidth]{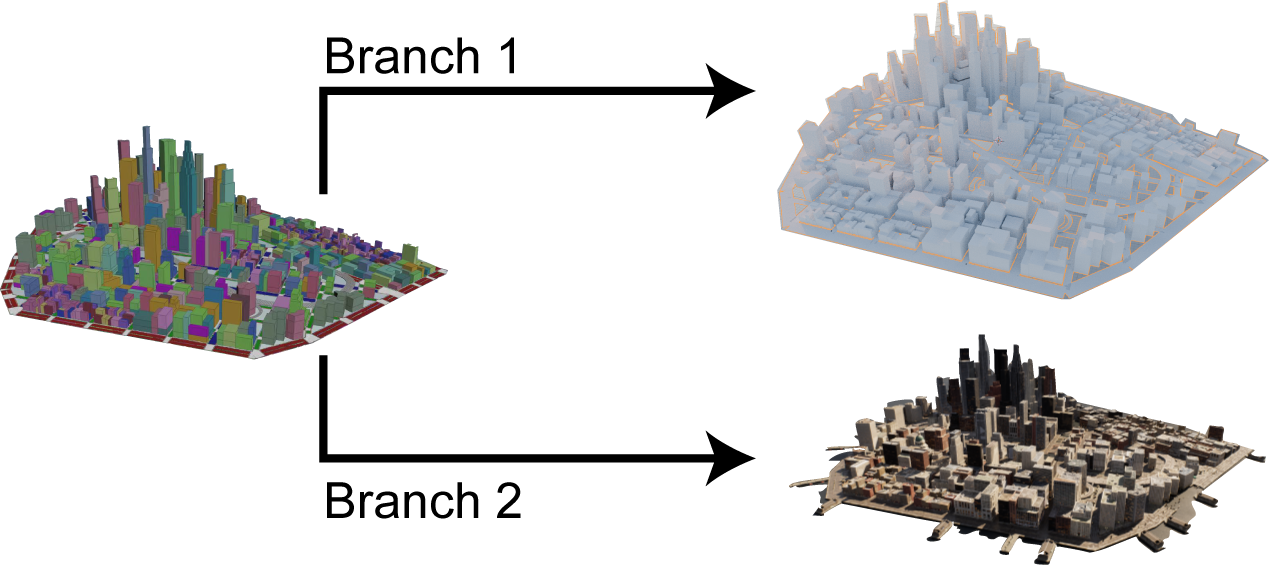}
    \caption{\textbf{Procedural Generation of Urban Environments.} This figure illustrates the two branches of the output from the PCG process. (a) shows the PCG output; (b) shows the 3D geometry (polygons) used for radio simulation; (c) shows the photorealistic rendering in UE5.}
    \label{fig:scene_compose}
\end{figure}

\subsection{VAS Reconstruction}

As identified in Section \ref{sec:motivation}, standard planar projections introduce a geometric disparity between the ego-centric visual domain and the allocentric radio domain. To bridge this gap, SynthRM implements a VAS reconstruction strategy. The primary objective of this process is to establish a rigorous, fine-grained geometric correspondence between the optical and radio modalities. This alignment is not merely a function of data resolution. Instead, it ensures that the physical anchor points for radio simulation are identical to the surfaces captured by the visual sensor. While the underlying 3D scene model contains the structural information for propagation, utilizing its sparse geometry directly would result in a misalignment with the continuous visual input. By performing an inverse rendering operation that lifts the 2.5D depth buffer into a 3D Euclidean manifold, we ensure the simulation is strictly congruent with the sensor's view frustum.

\nosubsection{Manifold Back-Projection}

Let the image domain be defined on a discrete lattice $\Omega \subset \mathbb{Z}^2$ of dimension $H \times W$. We are given a depth map $D: \Omega \to \mathbb{R}^+$ and a camera model parameterized by the intrinsic matrix $\mathbf{K}$ and the extrinsic transformation $\{\mathbf{R}, \mathbf{t}\}$. We define a lifting mapping $\Psi$ that transforms a pixel coordinate $\mathbf{u} \in \Omega$ into a vertex in the global world frame:
\begin{equation}
    \mathbf{p}_c(u,v) = D(u,v) \cdot \mathbf{K}^{-1} \tilde{\mathbf{u}}
\end{equation}
\begin{equation}
    \Psi(u,v) = \mathbf{R}^\top (\mathbf{p}_c(u,v) - \mathbf{t})
\end{equation}
By applying $\Psi$ over the entire domain, we generate the vertex set $\mathcal{V}$ of the visible surface mesh.

\nosubsection{Topological Triangulation}

To establish surface continuity for radio sampling, we construct a mesh $\mathcal{M}$ by inducing a topology over the vertex set. Exploiting the structured adjacency of the image lattice, we employ a deterministic triangulation scheme that decomposes each pixel quad into triangular faces. Crucially, this reconstructed mesh $\mathcal{M}$ operates exclusively as a \textit{receiver manifold}, which is a virtual surface that defines the spatial distribution of receiver probes, rather than as a physical obstacle affecting propagation. The ray-tracing engine calculates complex electromagnetic interactions based on the full high-fidelity scene geometry generated in Phase I, while $\mathcal{M}$ serves solely to discretize the spatial query points. This design ensures that the resulting radio map is sampled exactly at the locations where the optical sensor perceives the environment.

\begin{figure}[htbp]
    \centering
    \includegraphics[width=0.45\textwidth]{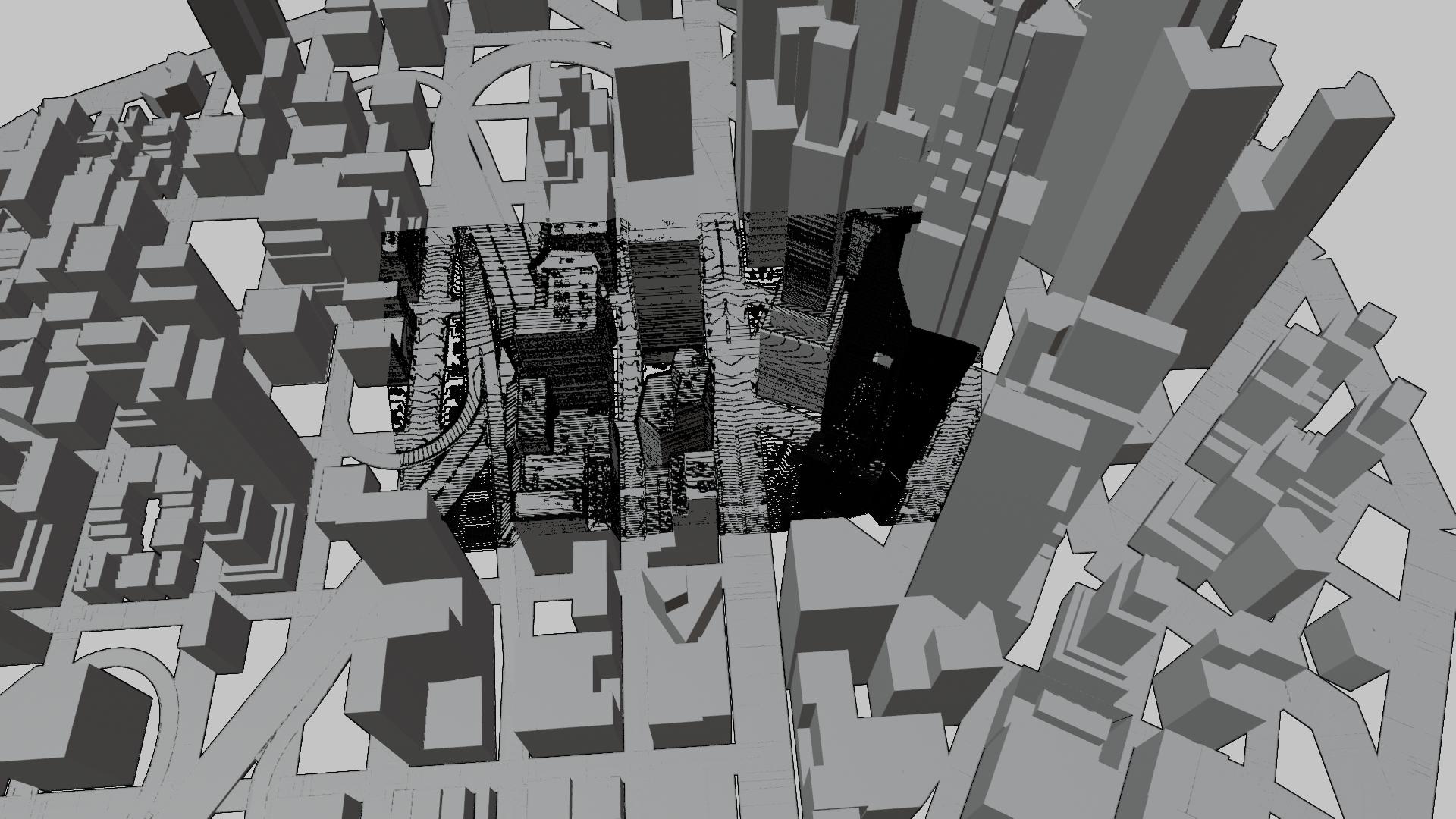}
    \caption{The reconstructed surface for a specific perspective camera overlayed on the 3D city scene model. Decimation of vertices has been performed during rendering to clearly demonstrate the 3D structure.}
    \label{fig:recon-surf}
\end{figure}

\subsection{Radio Simulation on VAS}
In the final phase, the reconstructed VAS mesh serves as the spatial anchor for electromagnetic simulation. To establish a precise correspondence between visual features and radio properties, we instantiate a discrete receiver (Rx) probe at the geometric centroid of each triangular face generated in the previous step.

This formulation addresses the critical density disparity between the scene's geometric definition and the optical sensor's resolution. The underlying 3D scene model, while structurally accurate for propagation, consists of sparse vertices and simplified meshes that lack the granularity for pixel-level alignment. Relying on this original geometry would yield non-uniformly distributed radio patches, severing the strict correspondence required for dense vision tasks. By defining the receiver manifold through depth back-projection, we generate a high-density lattice that matches the image resolution at pixel level. This ensures that every visual pixel allows the model to leverage fine-grained visual textures as semantic cues to identify dominant structural blockers.

\nosubsection{Antenna Pattern Modeling}

To rigorously capture the spatial characteristics of wireless propagation, we explicitly define the antenna properties for both link ends. The receiver is modeled as a single isotropic antenna with vertical polarization. For the transmitter, we employ an antenna pattern model defined by the receiving power measured on a sphere with unit radius. We implement three distinct configurations—SISO, $4\times4$ MIMO, and $8\times4$ MIMO—to represent varying degrees of spatial diversity and beamforming capability. These configurations yield distinct spatial energy distributions, determining how the transmitted energy interacts with the scene geometry before reaching the surface-bound receivers.

\begin{figure}[htbp]
    \centering
    \includegraphics[width=0.45\textwidth]{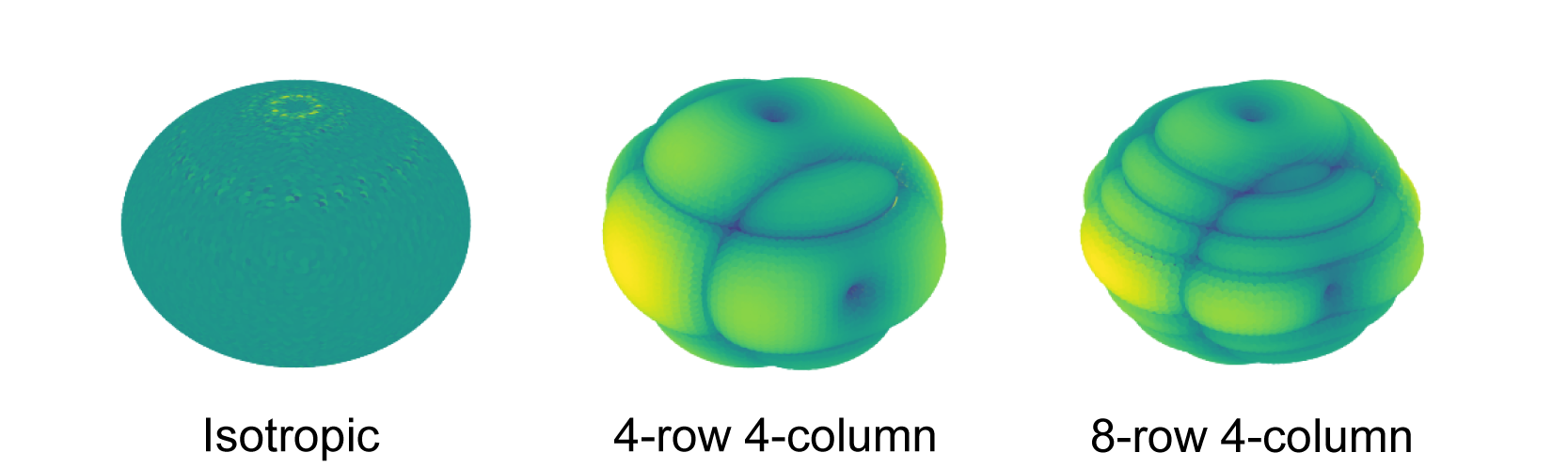}
    \caption{Radiation patterns for the three transmitter antenna configurations used in SynthRM. The SISO model exhibits uniform radiation, while the MIMO arrays demonstrate directional beamforming characteristics.}
    \label{fig:antenna}
\end{figure}

\nosubsection{Simulation Configuration}

The transmitter placement is designed to emulate ground-based mobile agents. Transmitters are sampled on a horizontal plane elevated 1.6 meters above the reference datum to ensure consistent clearance above the terrain. We enforce physical plausibility by culling any candidate Tx locations that fall within building footprints. With the Tx-Rx pairs defined, the solver computes the deterministic propagation paths to determine the signal quality at each mesh centroid. The resulting metrics are fused directly onto the visible geometry, producing a ``Radio-Textured Mesh'' that aligns strictly with the optical input.
\section{SynthRM Dataset}

Using SynthRM, we generate a large-scale dataset for vision-aided wireless sensing that captures UAV perspectives across procedural urban environments, featuring skyscraper-dense Downtown, transitional Mix, and peripheral Margin blocks. Unlike standard computer vision datasets that focus solely on optical consistency, or radio datasets that rely on sparse point measurements, SynthRM provides dense, pixel-aligned data cubes where visual geometry and electromagnetic properties are inextricably linked.

\subsection{Data Modalities and Annotations}
\begin{figure*}[htbp]
    \centering
    \begin{subfigure}{0.32\textwidth}
        \centering
        \includegraphics[width=\textwidth]{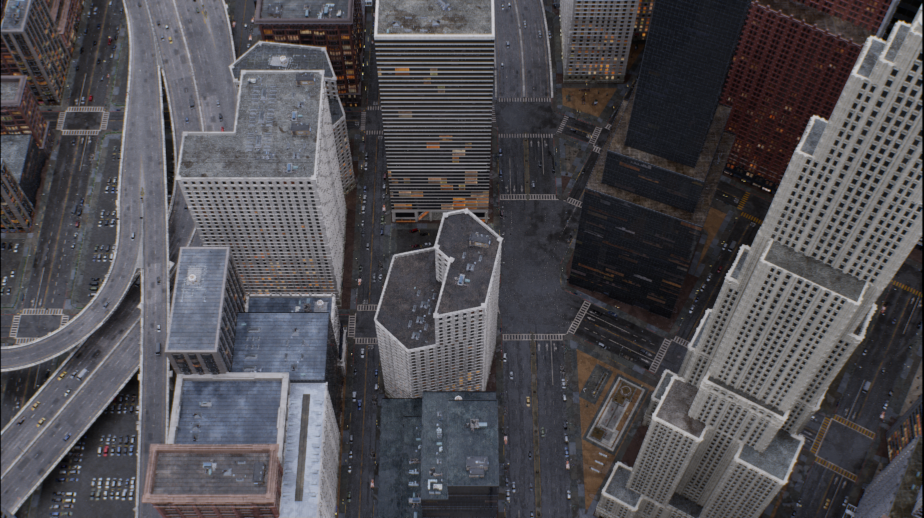} 
        \caption{RGB Image}
    \end{subfigure}
    \begin{subfigure}{0.32\textwidth}
        \centering
        \includegraphics[width=\textwidth]{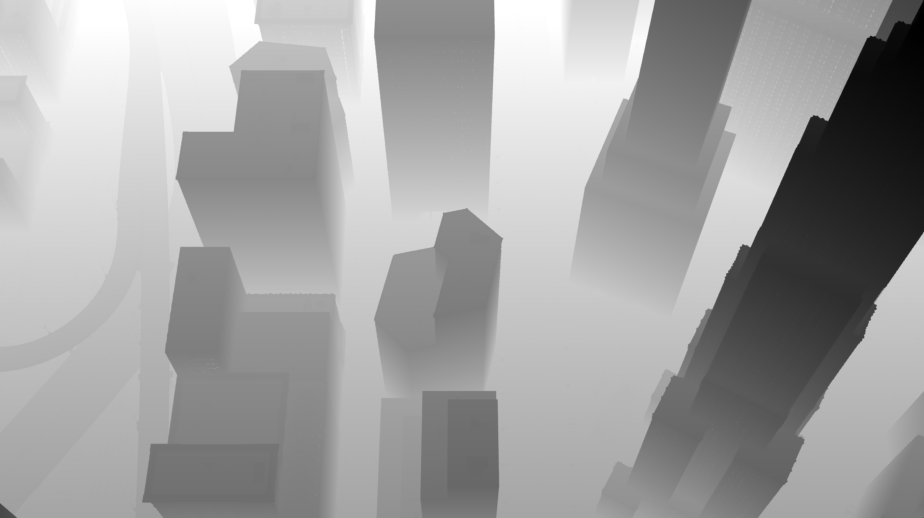} 
        \caption{Depth}
    \end{subfigure}
    \begin{subfigure}{0.32\textwidth}
        \centering
        \includegraphics[width=\textwidth]{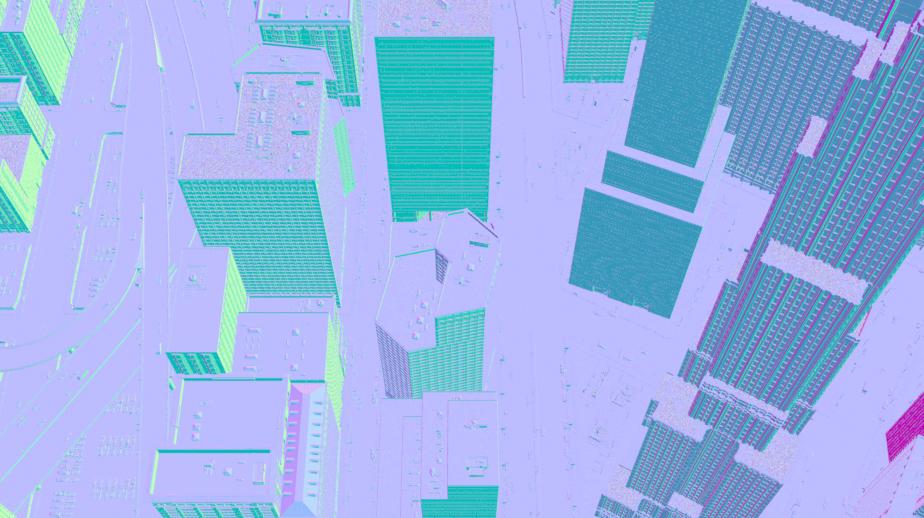} 
        \caption{Surface Normals}
    \end{subfigure}
    \begin{subfigure}{0.32\textwidth}
        \centering
        \includegraphics[width=\textwidth]{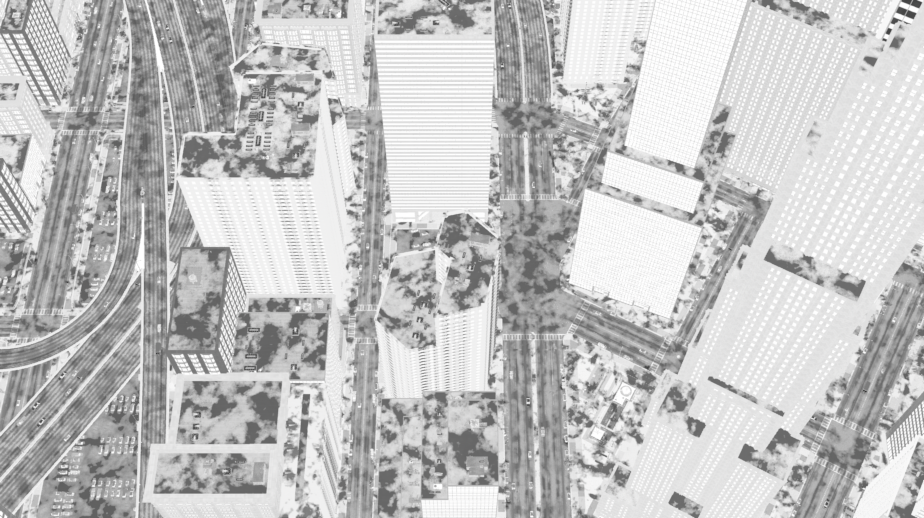} 
        \caption{Roughness}
    \end{subfigure}
    \begin{subfigure}{0.32\textwidth}
        \centering
        \includegraphics[width=\textwidth]{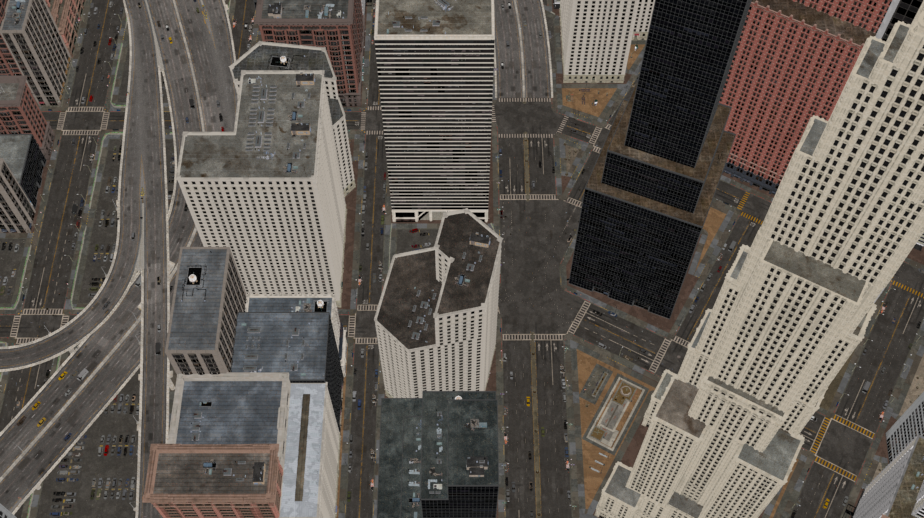} 
        \caption{Diffuse}
    \end{subfigure}
    \begin{subfigure}{0.32\textwidth}
        \centering
        \includegraphics[width=\textwidth]{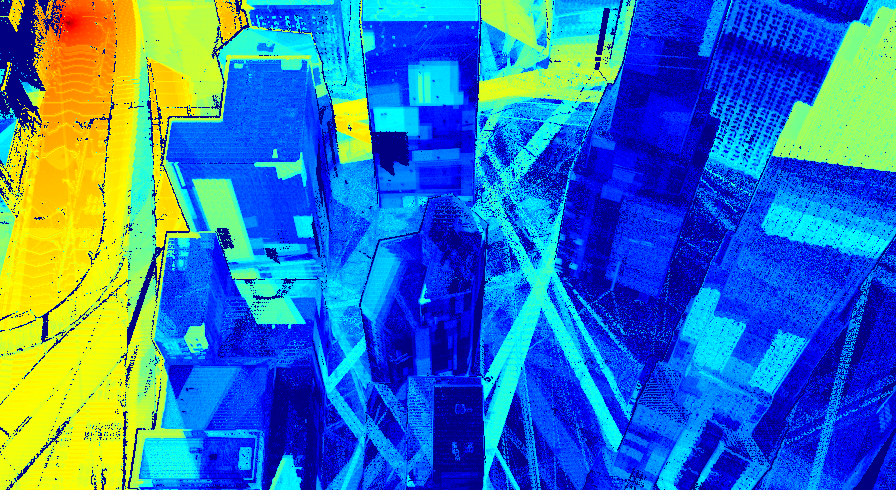} 
        \caption{Radio Map}
    \end{subfigure}
    \caption{\textbf{Multi-Modal Data Sample.} Every frame includes pixel-aligned channels: (a) RGB Image, (b) Depth Map, (c) Surface Normals, (d) Roughness, (e) Diffuse Material properties, and (f) The Radio Map projected onto the VAS.}
    \label{fig:modalities}
\end{figure*}

Each sample in the dataset is a synchronized tensor containing multiple aligned modalities. As illustrated in Figure~\ref{fig:modalities}, the primary optical inputs include high-dynamic-range RGB images and logarithmic depth maps. To facilitate physically grounded learning, we also export the underlying geometric and material buffers used during the rendering process. These include Surface Normals, which govern the angle of incidence, as well as Roughness maps and Diffuse material properties.

Crucially, the targeted radio map is mapped onto the same perspective grid, providing the Path Gain and SINR values for every visible surface fragment. This dense alignment allows models to learn the direct mapping from geometric properties to radio behavior. To foster generalization, the radio simulation employs a simplified material abstraction (categorizing surfaces primarily as concrete or ground). This design choice intentionally reduces the dependence on complex dielectric parameters, encouraging the learning model to focus on geometric determinism, specifically how structural shapes, edges, and incidence angles dictate signal attenuation, rather than overfitting to specific material textures.

\subsection{Environmental Variations}
\begin{figure}[htbp]
    \centering
    \begin{subfigure}{0.15\textwidth}
        \centering
        \includegraphics[width=\textwidth]{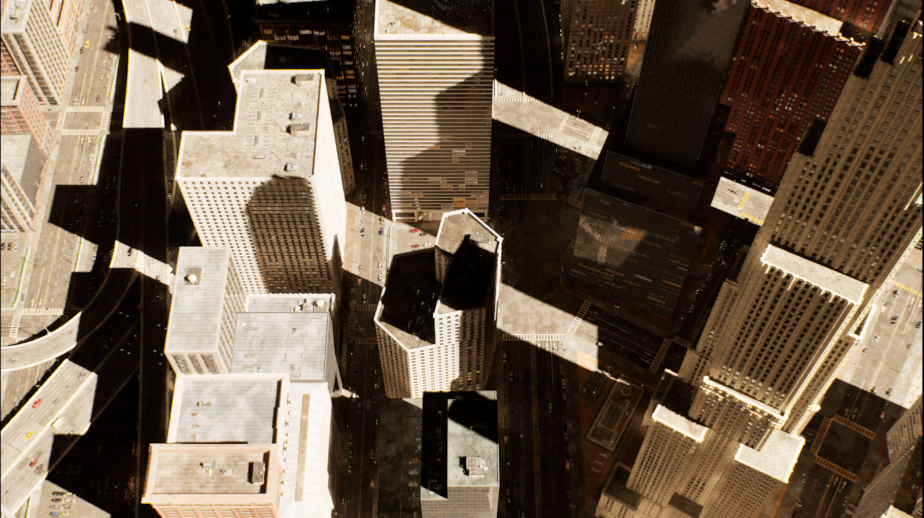}
        \caption{Sunny}
    \end{subfigure}
    \begin{subfigure}{0.15\textwidth}
        \centering
        \includegraphics[width=\textwidth]{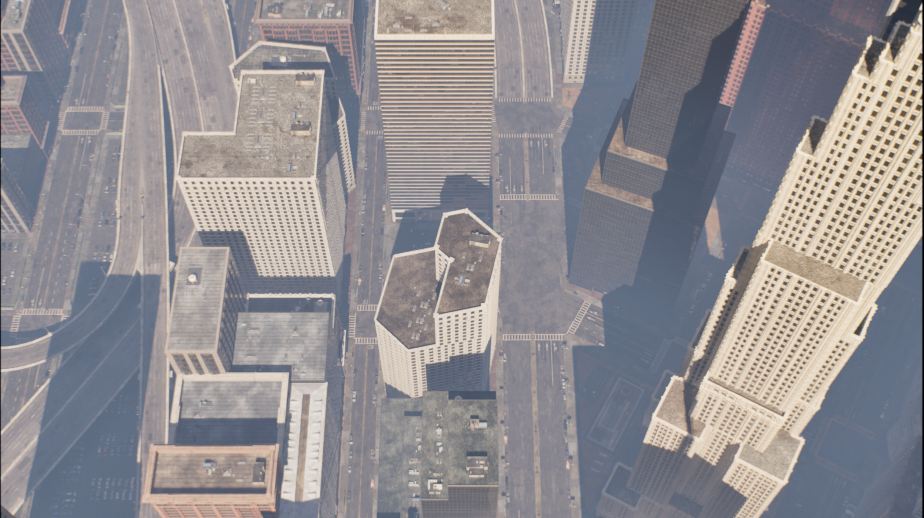}
        \caption{Foggy}
    \end{subfigure}
    \begin{subfigure}{0.15\textwidth}
        \centering
        \includegraphics[width=\textwidth]{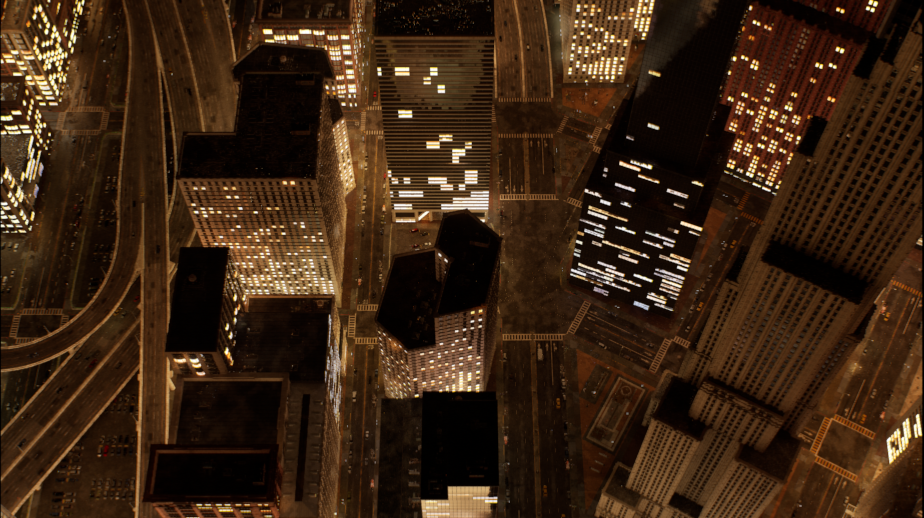}
        \caption{Night}
    \end{subfigure}
    \begin{subfigure}{0.47\textwidth}
        \centering
        \includegraphics[width=0.32\textwidth]{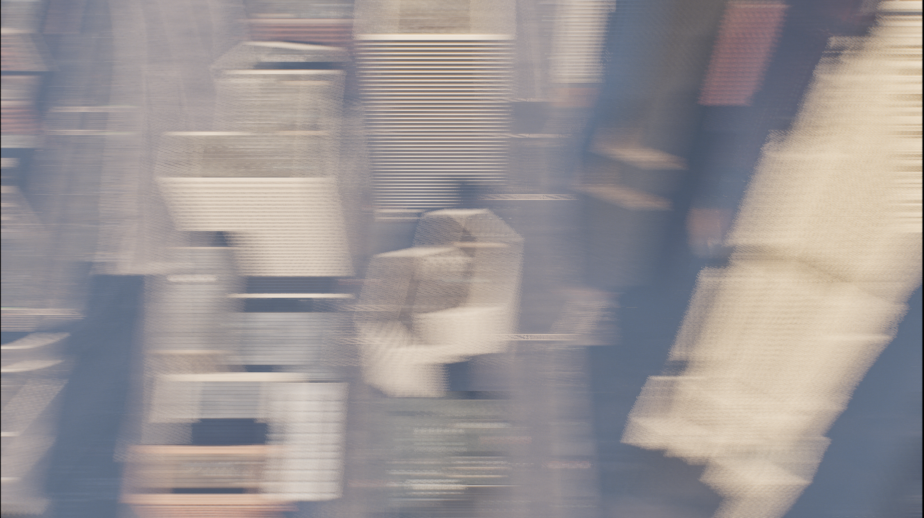}
        \includegraphics[width=0.32\textwidth]{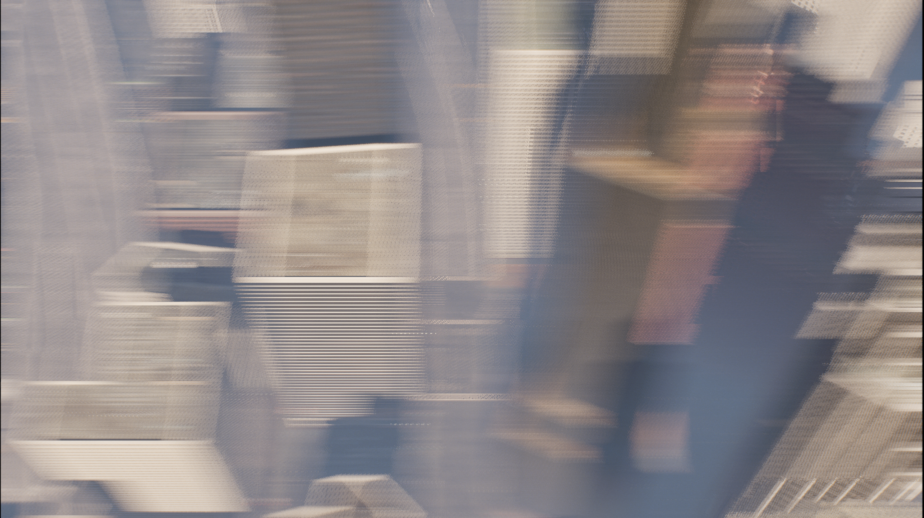}
        \includegraphics[width=0.32\textwidth]{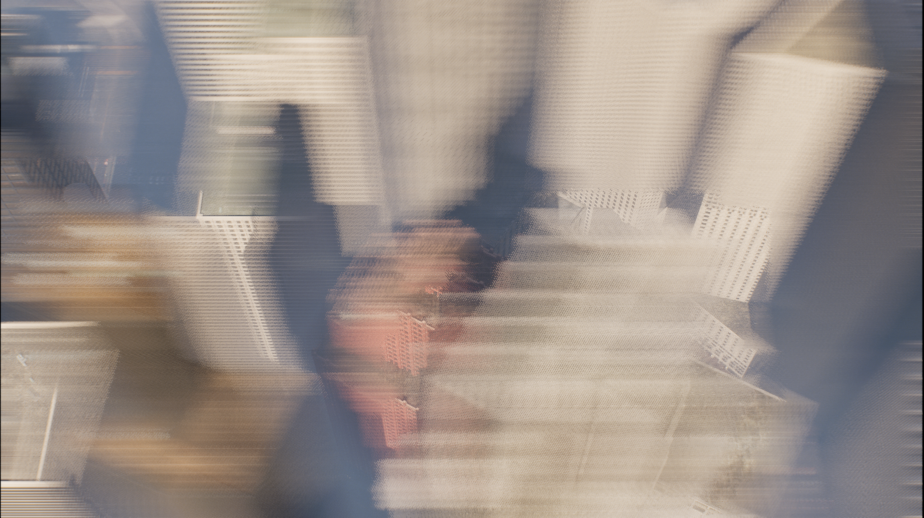}
        \caption{Motion Blur}
        \label{fig:motion_blur}
    \end{subfigure}
    \caption{\textbf{Variations.} The dataset includes diverse conditions to enhance model robustness.}
    \label{fig:env_cond}
\end{figure}

To ensure the trained models are robust to the domain shifts encountered in real-world deployment, the dataset incorporates significant environmental variability. As shown in Figure~\ref{fig:env_cond}, we procedurally alter the lighting and atmospheric conditions of the scenes. The dataset spans multiple times of day, ranging from high-noon clear skies to low-light nighttime scenarios, testing the model's ability to infer geometry when visual texture is degraded. Furthermore, we simulate adverse weather conditions such as fog, which alters the transmission of light without significantly affecting radio propagation frequencies below mmWave.

We also introduce camera artifacts like motion blur to mimic the characteristics of high-speed mobile agents such as UAVs. As shown in Figure~\ref{fig:motion_blur}, these variations are far less informative for visual-radio inference, but are actually common in real-world deployments. With such challenging conditions compel to the dataset, we aim to push the limits of vision-aided wireless sensing models towards practical applicability.

\subsection{Graph-Based Collaborative Perception Orchestration}
\label{subsec:orchestration}

Robust vision-aided sensing requires data that captures the full geometric manifold of the environment. To maximize the geometric information content of our dataset, we implement a graph-based orchestration strategy that optimizes the spatial distribution of sampling agents.

We model the potential sampling space as a networked sensing graph $\mathcal{G} = (\mathcal{V}, \mathcal{E})$, where nodes represent sensing poses and edges quantify visual overlap based on matched Scale-Invariant Feature Structure (SIFT) features obtained by Colmap \cite{colmapmvs2016,colmapsfm2016}. By partitioning this graph using community detection algorithms \cite{de2011generalized}, we identify distinct ``perception communities''—clusters of viewpoints that observe the same geometric features.

This structured approach enhances the geometric representation of the dataset in two critical ways:
\begin{enumerate}
    \item \textbf{Aggregating Complementary Geometry:} By clustering spatially correlated views into perception communities, we organize the data to provide multiple, varied perspectives of the same structures. This ensures that the model receives complementary geometric information which is critical for resolving depth ambiguities and achieving robust scene understanding.
    \item \textbf{Ensuring Global Completeness:} The orchestration forces the sampling distribution to cover all distinct geometric facets of the scene (e.g., opposing sides of a street canyon). This improves the geometrical representation and allows models to learn global spatial consistency from local ego-centric inputs.
\end{enumerate}

This graph-based approach transforms the raw trajectory data into a structured collaborative sensing campaign, validating the model's ability to integrate diverse perspectives.
\section{Evaluation and Analysis}

As the first platform to propose the VAS radio mapping paradigm, we establish a baseline evaluation on the SynthRM dataset. Our analysis focuses on three key dimensions: \textbf{Statistical Diversity}, ensuring the procedurally generated scenarios cover a wide range of propagation conditions; \textbf{Geometric-Semantic Consistency}, quantifying the causal link between visual objects (e.g., buildings, roads) and radio signal strength; and \textbf{Physical Plausibility}, verifying that the synthetic signals respect fundamental propagation laws relative to depth and surface orientation.

\subsection{Statistical Diversity}

\begin{figure}
    \centering
    \begin{subfigure}[b]{0.45\linewidth}
        \centering
        \includegraphics[width=\textwidth]{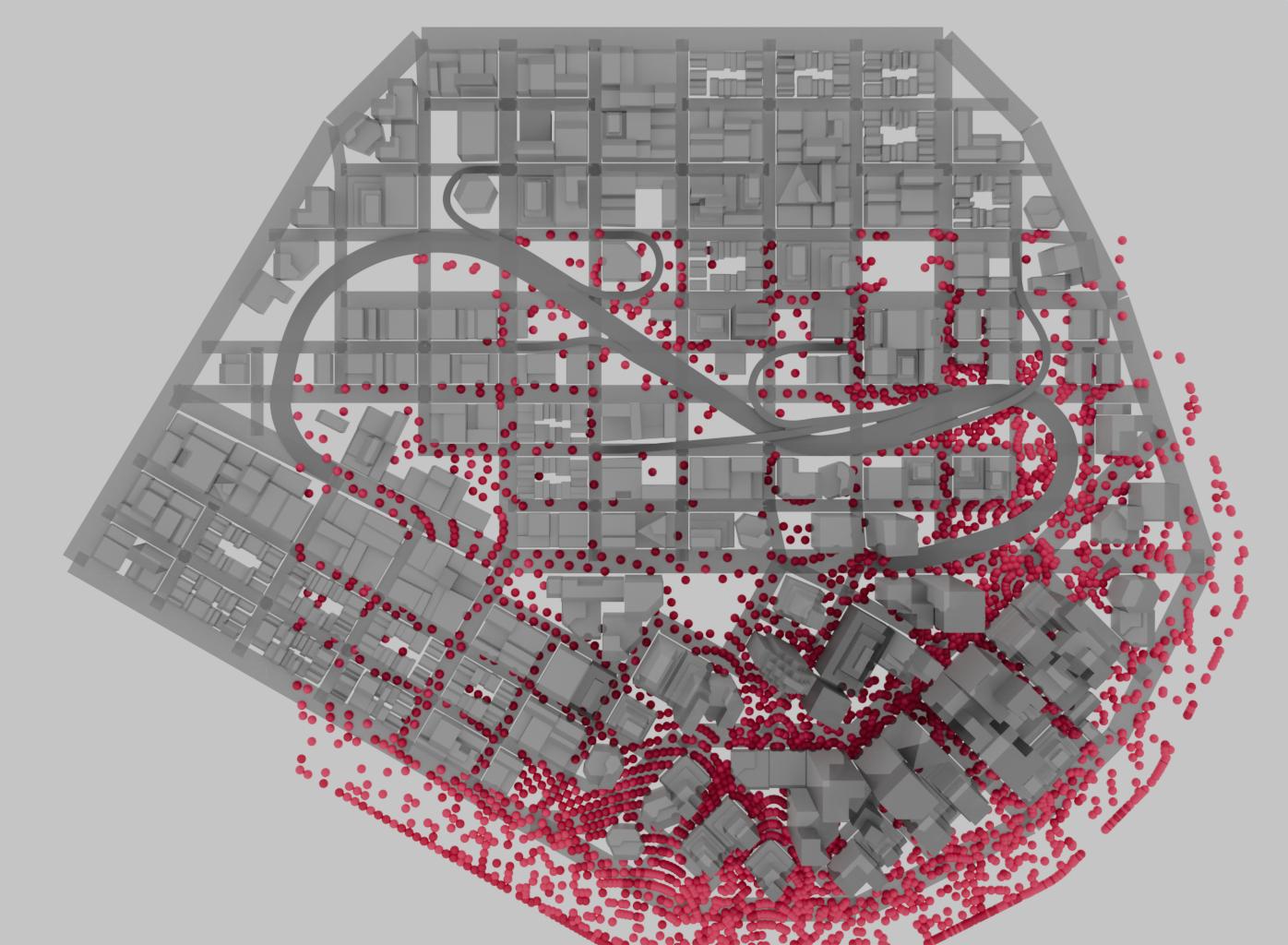}
    \end{subfigure}
    \begin{subfigure}[b]{0.45\linewidth}
        \centering
        \includegraphics[width=\textwidth]{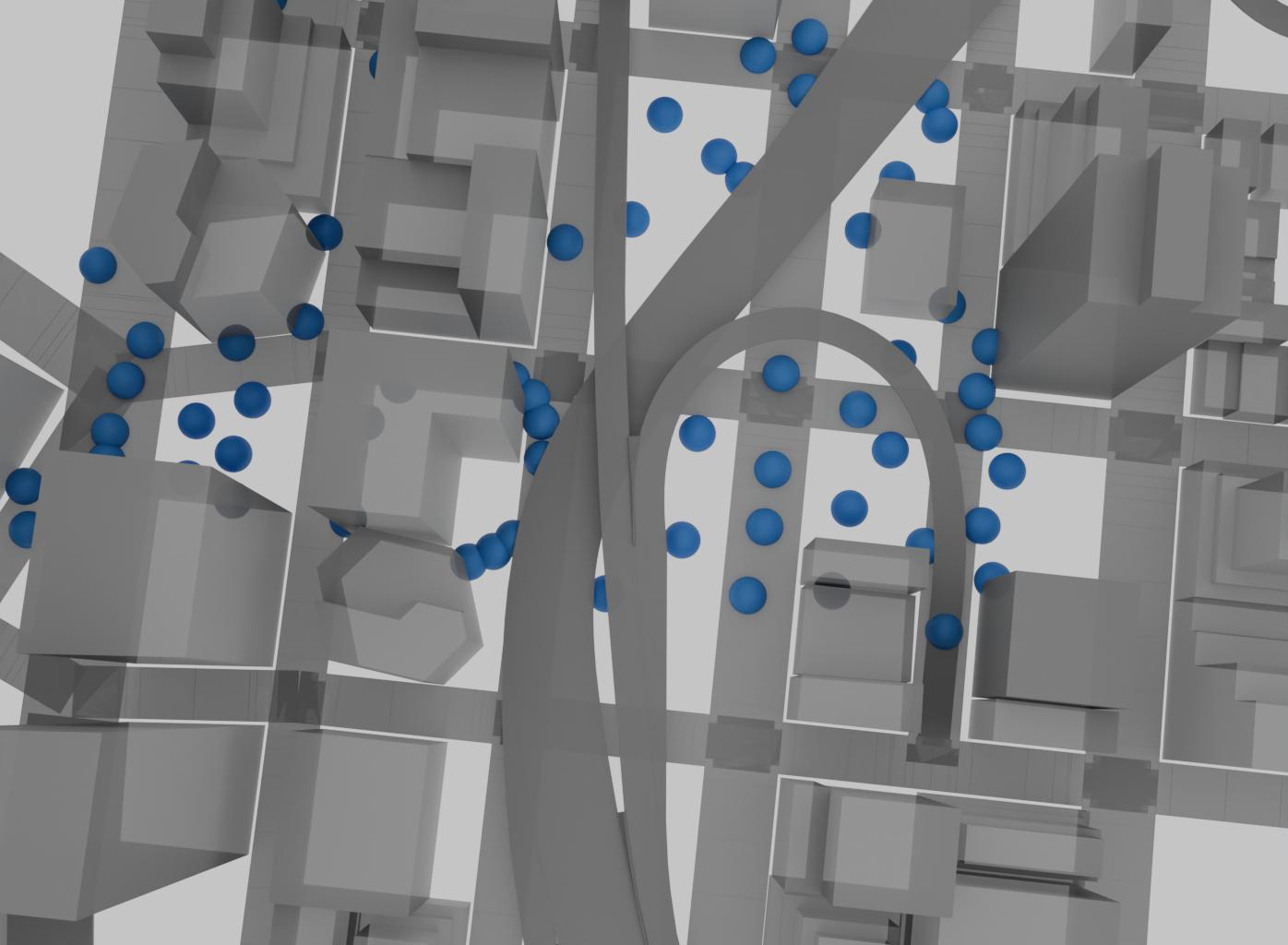}
    \end{subfigure}
    \caption{Transmitter position distribution. (a) illustrates the global perspective of transmitter positions within the Downtown area, while (b) provides a zoomed-in view of a specific urban block.}
    \label{fig:txpos}
\end{figure}

A robust dataset must exhibit high variance in both spatial topology and signal distribution to prevent model overfitting. 
We analyze the diversity of our dataset from two perspectives: the geometric layout of the environment and the stochastic properties of the signal propagation.

As shown in Figure~\ref{fig:txpos}, the procedurally generated transmitter positions exhibit high spatial entropy. 
By covering diverse semantic contexts, including open squares, narrow street canyons, and complex intersections, we ensure the dataset is not biased toward specific ``hotspot'' patterns. This spatial variance is critical for training generalizable models capable of handling the heterogeneous topologies found in real-world deployments.

\begin{figure*}[hbtp]
    \centering
    \includegraphics[width=\textwidth]{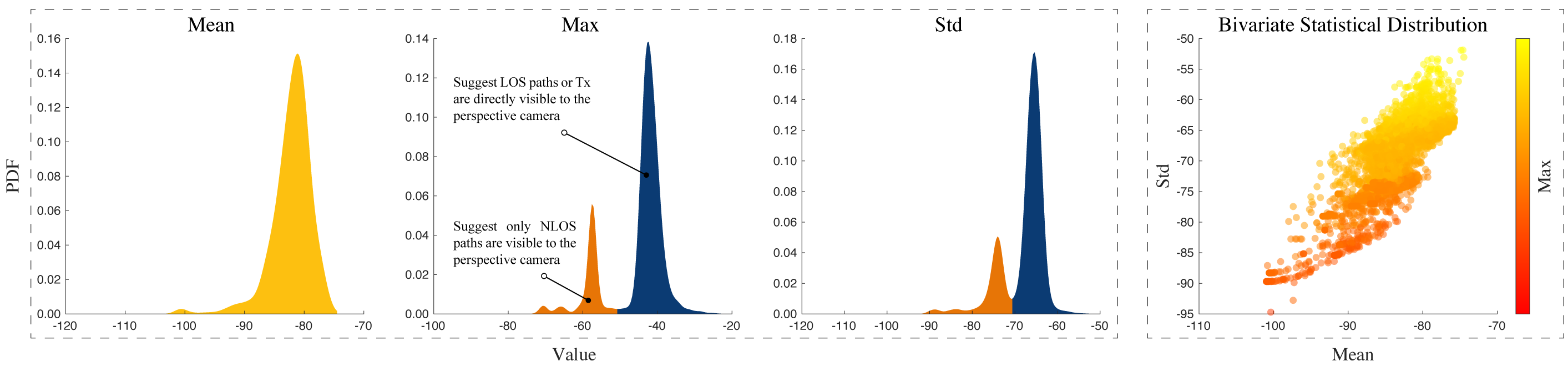}
    \caption{The statistics of signal diversity. The distinct bimodal peak in the Max Value distribution highlights the dataset's ability to distinguish between LoS and NLoS propagation regimes.}
    \label{fig:stat_dist}
\end{figure*}

For the signal distribution, we analyze the probabilistic characteristics of the path gains. We compute the mean values ($\mu$), maximum values ($\mathrm{P}_\mathrm{max}$), and standard deviations ($\sigma$) of path gains aggregated over distinct urban topologies. The probability distribution functions (PDFs) illustrated in Figure~\ref{fig:stat_dist} reveal a critical insight into the simulation's fidelity.

Specifically, the $\mathrm{P}_\mathrm{max}$ distribution exhibits a clear bimodal structure. This separation corresponds to the physical dichotomy of wireless channels: the high-power mode represents Line-of-Sight (LoS) dominance where transmitters are directly visible to the perspective camera, while the lower-power mode captures Non-Line-of-Sight (NLoS) scenarios driven purely by reflection and diffraction. To further visualize this diversity, the bivariate bubble plot in Figure~\ref{fig:stat_dist} maps the $\mu$ and $\sigma$ of each sample to a 2D manifold, with color intensity representing $\mathrm{P}_\mathrm{max}$. The wide dispersion of points confirms that SynthRM encompasses a continuous spectrum of propagation scenarios, effectively bridging the gap between high-strength, low-variance LoS links and deep, high-variance shadowing regions characteristic of urban fading channels.

\goodbreak
\subsection{The SynthRM Benchmark}

To formally quantify the utility of visual-aligned data, we introduce the \textbf{SynthRM Benchmark}. This evaluation framework serves a dual purpose: it measures the capability of visual semantics to serve as a valid proxy for radio signal inference, and it tests the feasibility of deployment in scenarios where explicit geometric models are unavailable.

\nosubsection{Baselines and Modifications}

Current data-driven radio mapping approaches typically rely on explicit geometric inputs like Digital Surface Models (DSMs) or height maps to predict coverage. However, in many real-world deployment scenarios, such high-fidelity topological data is either unavailable, outdated, or computationally expensive to maintain. A realistic vision-aided system must be capable of inferring the radio environment using only readily available sensor data.

To address this, we adapted three state-of-the-art architectures originally designed for top-down map prediction: RadioUNet \cite{radiounet2021}, PMNet \cite{pmnet2024}, and UNetDCN \cite{rmberlin2025}. We introduced a critical constraint to these baselines: geometric deprivation. We stripped the models of any explicit depth or height map inputs, forcing them to rely exclusively on two easy-to-acquire modalities:
\begin{enumerate}
    \item \textbf{Ego-centric RGB Images:} Providing implicit geometric cues and material semantics from the perspective of a standard camera.
    \item \textbf{Transmitter Metadata:} Consisting of the transmitter's 3D position (obtainable via GPS/positioning systems) and its antenna pattern (calculable via standard models).
\end{enumerate}

These modified architectures, denoted with the suffix '-S' (e.g., RadioUNet-S), are trained to predict the perspective-aligned radio map. This formulation validates a practical deployment pipeline: assuming a model is well-trained on SynthRM, an agent can infer complex electromagnetic coverage using only a video feed and basic network parameters, without requiring a pre-built digital twin of the city's topology.

\begin{table*}[htbp]
    \caption{Performance comparison of different models on SynthRM dataset across various urban blocks.}
\begin{tabularx}{\textwidth}{c|CCPP|CCPP|CCPP} 
\hline
\multirow{2}{*}{\textbf{Block}}       & \multicolumn{4}{c|}{\textbf{RadioUNet-S}}                                                 & \multicolumn{4}{c|}{\textbf{PMNet-S}}                                                     & \multicolumn{4}{c}{\textbf{UNetDCN-S}}                               \\
                                      & NMSE$\downarrow$ & MAE$\downarrow$ & PSNR$\uparrow$ & \multicolumn{1}{c|}{SSIM$\uparrow$} & NMSE$\downarrow$ & MAE$\downarrow$ & PSNR$\uparrow$ & \multicolumn{1}{c|}{SSIM$\uparrow$} & NMSE$\downarrow$ & MAE$\downarrow$ & PSNR$\uparrow$ & SSIM$\uparrow$ \\ \hline
\textit{\underline{Mix}} & 0.7604           & 0.1346          & 15.31          & 0.43                                & 0.9844           & 0.1474          & 14.21          & 0.36                                & 0.7803           & 0.1335          & 15.21          & 0.43           \\
Margin                                & 1.1208           & 0.1734          & 13.51          & 0.06                                & 133.8370         & 1.6605          & -7.22          & 0.02                                & 10.8383          & 0.4626          & 3.69           & 0.06           \\
Downtown                              & 1.2428           & 0.1794          & 13.04          & 0.08                                & 120.7216         & 1.5140          & -6.79          & 0.03                                & 12.3295          & 0.5110          & 3.11           & 0.08           \\ \hline
\end{tabularx}
\raggedright\footnotesize{\textit{Note: Data gathered from the Mix block were used for training these models.}}
\label{tab:model_performance}
\end{table*}

\nosubsection{Applicability Analysis}

The quantitative results in Table~\ref{tab:model_performance} provide empirical validation of the SynthRM platform's applicability for deep learning workflows. First, the convergence of all three baseline architectures within the training domain, where each achieves a Normalized Mean Squared Error (NMSE) of less than 1.0, demonstrates the fundamental feasibility of vision-aided radio mapping. This confirms that the perspective-aligned data carries sufficient causal information for neural networks to learn the mapping from visual semantics to radiometric responses.

Comparatively, RadioUNet-S exhibits the most robust performance, maintaining the lowest error rates and highest structural similarity (SSIM) across all topologies. However, the substantial performance drop observed in PMNet-S and UNetDCN-S when applied to unseen environments (\textit{Downtown}, \textit{Margin}) underscores the inherent complexity of the VAS domain. These degradation patterns reveal that legacy architectures, originally optimized for translation-invariant top-down maps, struggle to generalize when geometric scale varies with perspective depth. Rather than indicating a dataset defect, this ``generalization gap'' highlights the rigorous nature of the SynthRM benchmark, motivating the future development of specialized 3D-aware architectures capable of handling projective geometry.

\goodbreak
\nosubsection{Semantic-Physical Metrics}

To rigorously assess the correlation between specific visual objects and radio behavior, we employ the Point-Biserial Correlation Coefficient ($r_{pb}$). This metric quantifies the relationship between a binary variable (the presence of a specific semantic concept) and a continuous variable (radio signal strength). For a given semantic concept $C$ (e.g., ``Building''), let the segmentation mask $M \in \{0, 1\}^{H \times W}$ be the binary indicator where $M_{uv}=1$ if pixel $(u,v)$ belongs to class $C$, and $0$ otherwise. Let $S \in \mathbb{R}^{H \times W}$ be the corresponding aligned path gain map. The point-biserial correlation is calculated as:
\begin{equation}
    r_{pb} = \frac{\mu_1 - \mu_0}{\sigma_S} \sqrt{\frac{n_1 n_0}{n^2}}
\end{equation}
where $\mu_1$ and $\mu_0$ are the mean signal strengths of pixels belonging to the concept and the background, respectively; $\sigma_S$ is the standard deviation of the entire signal map; and $n_1, n_0$ are the pixel counts. A high positive $r_{pb}$ indicates that the visual presence of a concept strongly predicts radio energy.

\subsection{Benchmark Analysis}
We utilize the Segment Anything Model 3 (SAM3) \cite{carion2025sam3segmentconcepts} to automatically annotate the RGB images with open-vocabulary semantic masks, allowing us to calculate $r_{pb}$ for key urban concepts across varying district topologies.

\nosubsection{The Dominance of Verticality} 

As detailed in Table~\ref{tab:concepts}, the concept \textit{Buildings} exhibits the highest correlation ($r_{pb} \approx 0.8$) in \textit{Downtown} districts. This empirically demonstrates the \textit{Urban} \textit{Canyon} effect, where vertical facades act as the primary drivers of signal distribution by confining energy through multipath reflections. This finding validates the fundamental premise of our VAS approach: a traditional top-down map would inherently discard these vertical interactions, whereas SynthRM explicitly captures the surface-to-signal causality.

\nosubsection{Environmental Sensitivity} 

\begin{figure*}[htbp]
    \centering
    \includegraphics[width=\textwidth]{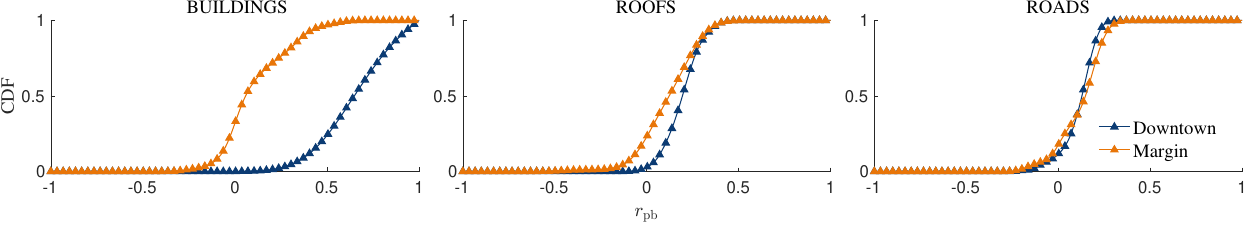}
    \caption{The CDF of the point-biserial correlation coefficients between the presence of buildings and radio signal strength. The distinct separation between \textit{Downtown} and \textit{Margin} validates the simulator's sensitivity to urban density.}
    \label{fig:biserial}
\end{figure*}

The cumulative distribution function (CDF) of $r_{pb}$ values, presented in Figure~\ref{fig:biserial}, reveals a distinct topological shift between \textit{Downtown} and \textit{Margin}. The steeper rise and right-shift of the \textit{Downtown} curve for ``\textit{Buildings}'' indicate that semantic geometry is a far more critical predictor of radio coverage in dense environments than in open suburbs. Conversely, ``\textit{Roads}'' show consistently lower correlation across all environments. This suggests that while ground reflections contribute to the channel, they are less discriminative than the blocking and reflecting effects of large structures. These results confirm that SynthRM captures the nuanced physics-semantics relationship: radio maps are not random textures, but are causally tethered to the 3D semantic geometry of the scene.

\nosubsection{Ecosystem Compatibility and Applicability} 

Beyond validating physical causality, these correlations demonstrate the immediate compatibility of SynthRM with the broader computer vision ecosystem. The strong linkage between standard visual semantics and radio signal attenuation indicates that off-the-shelf vision backbones can be effectively repurposed for wireless inference without architectural overhaul. This compatibility effectively addresses the ``Cold Start'' problem in digital twin deployment: rather than requiring expensive, time-consuming drive-test campaigns to build a radio map, a SynthRM-trained agent can infer the electromagnetic environment instantaneously from a single RGB-D frame. This capability is essential for latency-critical 6G applications, such as proactive handover and holographic beamforming \cite{bi2019engineering, romero2022radio}, where the computational cost of runtime ray-tracing is prohibitive.

\begin{table*}[htbp]
\centering
\begin{tabular}{>{\centering\arraybackslash}p{0.23\textwidth}>{\centering\arraybackslash}p{0.23\textwidth}>{\centering\arraybackslash}p{0.23\textwidth}>{\centering\arraybackslash}p{0.23\textwidth}} 
\toprule
\multirow{2}{*}{\textbf{Concept}} & \multicolumn{3}{c}{\textbf{Correlation} \textbf{Coefficient} (mean/std)}                        \\
                          & Downtown            & Margin         &          Mixed    \\ 
\midrule
Roads                     & 0.1168/0.0883       & 0.1236/0.1150  & 0.2344/0.0314    \\
Buildings                 & 0.7964/0.3043       & 0.1102/0.1766  & 0.5044/0.0867    \\
Roofs                     & 0.1961/0.0928       & 0.1192/0.1376  & 0.2448/0.0499    \\
Concrete                  & 0.0827/0.0896       & 0.0965/0.1830  & 0.0097/0.0260    \\
\bottomrule
\end{tabular}
\caption{Selected concepts and their correlation coefficients with radio signal strength in different urban blocks.}
\label{tab:concepts}
\end{table*}

\subsection{Case Study: Dynamic Urban Canyon Navigation}

\begin{figure}[htbp]
    \centering
    \includegraphics[width=0.45\textwidth]{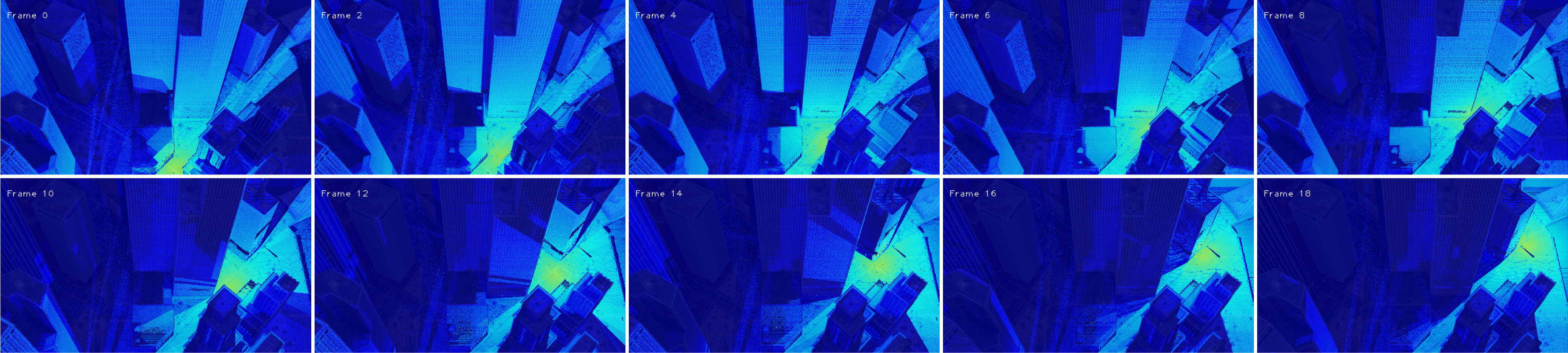}
    \caption{Case study illustrate a scenario where a vehicle is serving as a mobile base station, the SynthRM platform captures the dynamic radio environment and its interaction with urban geometry.}
    \label{fig:case_study}
\end{figure}

To contextualize the advantages of VAS in dynamic scenarios, we present a case study in Figure~\ref{fig:case_study}. Here, a vehicle equipped with a camera serves as a mobile base station, navigating through a dense urban canyon before emerging into an open area. This scenario serves as a stress test for geometric fidelity. Legacy top-down simulators and the deep learning models trained on them \cite{radiounet2021,pmnet2024,rmberlin2025} operate on the assumption of vertical invariance—collapsing a 3D building into a 2D footprint. Consequently, their predictions would be severely compromised in this environment, as they cannot distinguish between signal reception at street level versus the rooftop.

In contrast, SynthRM utilizes the VAS paradigm to accurately model signal propagation in full 3D space. The heatmaps in Figure~\ref{fig:case_study} reveal a phenomenon invisible to 2D representations: elevation-dependent signal distribution. As the vehicle moves, the upper strata of high-rise structures are distinctively ``lit up,'' indicating strong LoS connectivity, while the lower street-level facades often remain in deep shadow due to local occlusion. Furthermore, surfaces that are not directly exposed to the transmitting vehicle still exhibit significant receiving path gain. This captures the comprehensive result of multipath propagation, specifically the constructive interference caused by signal reflection, diffraction, and refraction off surrounding geometry.

This case study highlights the critical role of vertical geometry in mobile sensing. By resolving these complex 3D propagation paths, SynthRM empowers mobile systems to predict connectivity with a degree of reliability that 2D-topology models cannot achieve.

\section{Discussion}

The introduction of SynthRM represents a paradigm shift from loose, map-based heuristics to precise, pixel-aligned physical simulation. However, the deployment of such synthetic platforms in real-world mobile systems necessitates a critical examination of the domain gap, inherent limitations, and the prospective trajectory of this research.

\subsection{Necessity and Compatibility of SynthRM}

\begin{qbox}
    \textbf{Q1:} Current deep learning methods can already predict coverage using 2D maps or visual features. Why is the SynthRM platform necessary?
\end{qbox}

While existing deep learning approaches can extract features from 2D topologies, they fundamentally address an ill-posed problem: trying to infer complex 3D propagation from flat representations that lack vertical causality. SynthRM is necessary because it transforms this into a well-posed problem. By utilizing the VAS paradigm, we generate data that explicitly captures the verticality and 3D interactions missing from standard datasets.

Furthermore, SynthRM is not merely a tool for optimization, but a scalable platform designed to empower mobile computing systems. It enables the fast synthesis of massive-scale training data required to build robust neural networks. These models, trained on our physically aligned data, can be deployed directly on mobile edge devices such as autonomous vehicles, UAVs, and intelligent CCTV cameras to perform task-oriented work. Since our platform ensures that the input data matches what is physically accessible to a mobile agent, it bridges the gap between theoretical coverage prediction and practical, embodied deployment in dynamic urban environments.

\begin{qbox}
    \textbf{Q2:} Can a simulation-trained model truly generalize to the noisy real world?
\end{qbox}

We posit that SynthRM is best utilized not to replace real-world data, but to serve as a massive-scale \textbf{structural prior}. Deep learning models often struggle to learn high-frequency geometric dependencies from noisy real-world samples. SynthRM provides an idealized environment where these complex geometric-radio relationships are isolated and learnable. By instilling a strong geometric intuition during pre-training, models can subsequently be fine-tuned on smaller, noisy real-world datasets to adapt to specific sensor characteristics, bridging the sim-to-real gap more effectively than training on real data alone.

\subsection{Limitations}

Despite the high fidelity of our pipeline, we acknowledge two primary trade-offs. First, the accuracy of the radio simulation is bounded by the granularity of material definitions. Our simplified abstraction captures geometric determinism but introduces a material gap regarding complex dielectric properties found in the real world. Second, the VAS paradigm imposes a trade-off between spatial resolution and channel granularity. To achieve pixel-perfect spatial alignment for vision tasks, we utilize a coverage solver that outputs aggregated metrics rather than full Channel Impulse Responses (CIR), as generating full multipath profiles for millions of pixels remains computationally prohibitive for city-scale generation.

\subsection{Future Directions}
The roadmap for SynthRM extends beyond data generation into the realm of applied mobile systems, particularly where visual perception and wireless communication converge. Future work will focus on enhancing the diversity of synthetic environments while simultaneously deploying these models in critical infrastructure and autonomous systems.

\nosubsection{Parallel Systems for V2X Infrastructure} The platform is inherently compatible with the design of next-generation transportation infrastructure. By equipping smart street poles with both cameras and wireless access points, SynthRM can serve as a \textit{parallel system} or digital twin for Vehicle-to-Everything (V2X) networks \cite{ali2020passive, xu2022computer}. In this context, models trained on our platform can be deployed on edge nodes to infer the distribution of sensing and communication signals transmitted by vehicles in real-time. This allows the infrastructure to act as a proactive manager, predicting channel congestion or blockage based on traffic flow visibility before packet loss occurs.

\nosubsection{Embodied Intelligence in Wireless Swarms} As mobile agents such as drones and robotic clusters become more autonomous, the reliance on reactive network metrics (e.g., Round-Trip Time or packet loss rates) becomes a bottleneck. SynthRM enables a shift toward embodied intelligence where agents can proactively assess link quality. By using the proposed VAS data, a robot can predict path gains and SINR heatmaps directly from its visual feed. This allows self-organizing networks to optimize their topology based on predicted connectivity rather than reacting to signal degradation, ensuring robust communication in complex, obstacle-rich environments.

\nosubsection{Predictive QoS for Mobile XR} Finally, the platform holds significant promise for Mobile Augmented and Extended Reality (AR/XR). These applications demand consistently high throughput and ultra-low latency, making them highly sensitive to even momentary signal drops. SynthRM enables the development of \textit{Predictive Quality of Service} (QoS) mechanisms. By correlating the visual scene with radio propagation, an XR headset could anticipate ``radio shadows,'' areas of weak signal caused by static or dynamic occluders, before the user physically enters them. This look-ahead capability allows the system to aggressively pre-buffer high-fidelity holographic content or initiate seamless handover protocols, maintaining the illusion of immersion where reactive systems would stall.
\section{Related Work}

\label{sec:rel-work}

The integration of visual and radio-frequency modalities is critical for 6G environment-aware communication. However, a significant gap persists between available static datasets and the need for dynamic, physically aligned simulation platforms. This section reviews existing data resources and the emerging role of digital twin automation in bridging this gap.

\nosection{Radio Map and Multimodal Datasets}
Existing research has established baselines for data-driven wireless sensing, though often with geometric limitations. Channel-centric datasets, such as CKMImageNet \cite{ckmimagenet2024} and DeepMIMO \cite{deepmimo2019}, provide massive CSI matrices and channel knowledge maps. However, these are intrinsically coordinate-centric lookup tables lacking authentic optical modalities, which prevents models from learning the texture-to-channel mappings required for vision-aided sensing.

To incorporate geometry, morphological datasets like RadioMapSeer \cite{radiomapseer2022}, 3DRadioMap \cite{radiomapseer3d2022}, and PMNet \cite{pmnet2024} utilize building footprints or 2.5D height maps to predict radio coverage. While they demonstrate that neural networks can learn diffraction patterns, they rely on top-down orthographic projections. This ``bird's-eye'' view discards the vertical occlusion details inherent to the ego-centric perspective of ground-based agents.

Recent visual-enabled datasets attempt to bridge this modality gap. RMDirectionalBerlin \cite{rmberlin2025} and ViWi \cite{viwi2020} introduce high-resolution imagery and co-located wireless parameters. However, they often suffer from domain gaps (aerial vs. ground views) or loose coupling between visual rendering and radio simulation. Advanced multi-modal efforts like M$^3$SC \cite{m3sc2023}, SynthSoM \cite{bai2025multi, cheng2025synthsom} and the real-world DeepSense 6G \cite{deepsense6g2023} offer rich sensor suites (LiDAR, RGB, Radar). Yet, real-world data is limited by trajectory sparsity, while synthetic alternatives often mismatch ego-centric images with allocentric heatmaps, rendering cross-view inference ill-posed.

\nosection{Toward Digital Twin Simulation Platforms} 
High-fidelity digital twin networks are replacing static simulators to support 6G autonomous networks, enabling continuous synchronization and ``virtual drive-testing'' for mobility management \cite{lin20236g}. To address scalability in these dynamic environments, recent research leverages Generative AI to synthesize time-variant channel models \cite{tao2024wireless} and utilizes AutoML to automatically generate ``Unit Twins'' for network software validation \cite{agdt2025}.

The scope of digital twin networks has further expanded to encompass 3D mobility, from managing the extreme dynamics of non-terrestrial networks \cite{al2023digital} to supporting embodied agents via physics-informed reinforcement learning for indoor navigation \cite{li2025digital}. However, while recent frameworks have advanced regarding lifecycle management and resource optimization \cite{zhang2025digital}, they often lack the precise, ego-centric geometric alignment required to fully enable vision-aided sensing for mobile agents---a gap this work aims to bridge.
\section{Conclusion}

We present SynthRM, a scalable synthetic data platform that establishes a new standard for physically consistent vision-aided wireless sensing. By introducing the VAS simulation paradigm, we resolve the geometric disconnect inherent in prior methodologies, transforming cross-modal radio inference from an ill-posed ambiguity into a well-posed physical problem. Beyond serving as a static repository, SynthRM functions as a democratized engine for embodied intelligence, enabling researchers to generate diverse, city-scale training environments on consumer-grade hardware. The resulting dataset provides the dense, pixel-aligned structural priors necessary to bridge the sim-to-real gap. Ultimately, this work empowers next-generation mobile agents, from UAVs to autonomous vehicles, to move beyond reactive optimization and achieve proactive, geometry-aware connectivity in the complex landscape of 6G.
\begin{acks}
We thank LMark for his mitigation of MatrixCity Plugins from UE 5.0.3 to UE 5.5.4. \grant
\end{acks}

\bibliographystyle{ACM-Reference-Format}
\bibliography{base}
\appendix

\section{Implementation Details}

\subsection*{Simulation Details}
To ensure simulation fidelity, the city scene model is segmented into two material categories based on Sionna-RT's ITU radio material definitions~\cite{ituradio2023}. The material assignments are configured as follows:
\textbf{Buildings:} Assigned the \underline{concrete} material type.
\textbf{Terrain (Ground \& Roads):} Assigned the \underline{very dry ground} material type.
The detailed configuration for the ray-tracing simulation is provided in Table~\ref{tab:sim-params}.

\begin{table}[htbp]
\centering
\caption{Ray-Tracing Simulation Parameters}
\begin{tabularx}{\linewidth}{CC} 
\toprule
\multicolumn{1}{c}{\textbf{Parameter}} & \multicolumn{1}{c}{\textbf{Value}}  \\ 
\midrule
Frequency                     & $3.5\times 10^9$ Hz        \\
Bandwidth                     & $1\times 10^6$ Hz          \\
Power                         & 30dBm                      \\
Temperature                   & 293K                       \\
Max Mepth                     & 20                         \\
Sample per Tx                 & $2^{31}$                   \\
LoS                           & \cmark                     \\
Specular Reflection           & \cmark                     \\
Diffuse Reflection            & \xmark                     \\
Refraction                    & \cmark                     \\
Diffraction                   & \cmark                     \\
Edge Diffraction              & \xmark                     \\
Diffraction Lit Region        & \cmark                     \\
\bottomrule
\label{tab:sim-params}
\end{tabularx}
\end{table}

\begin{figure*}[htbp]
    \centering
    \includegraphics[width=\textwidth]{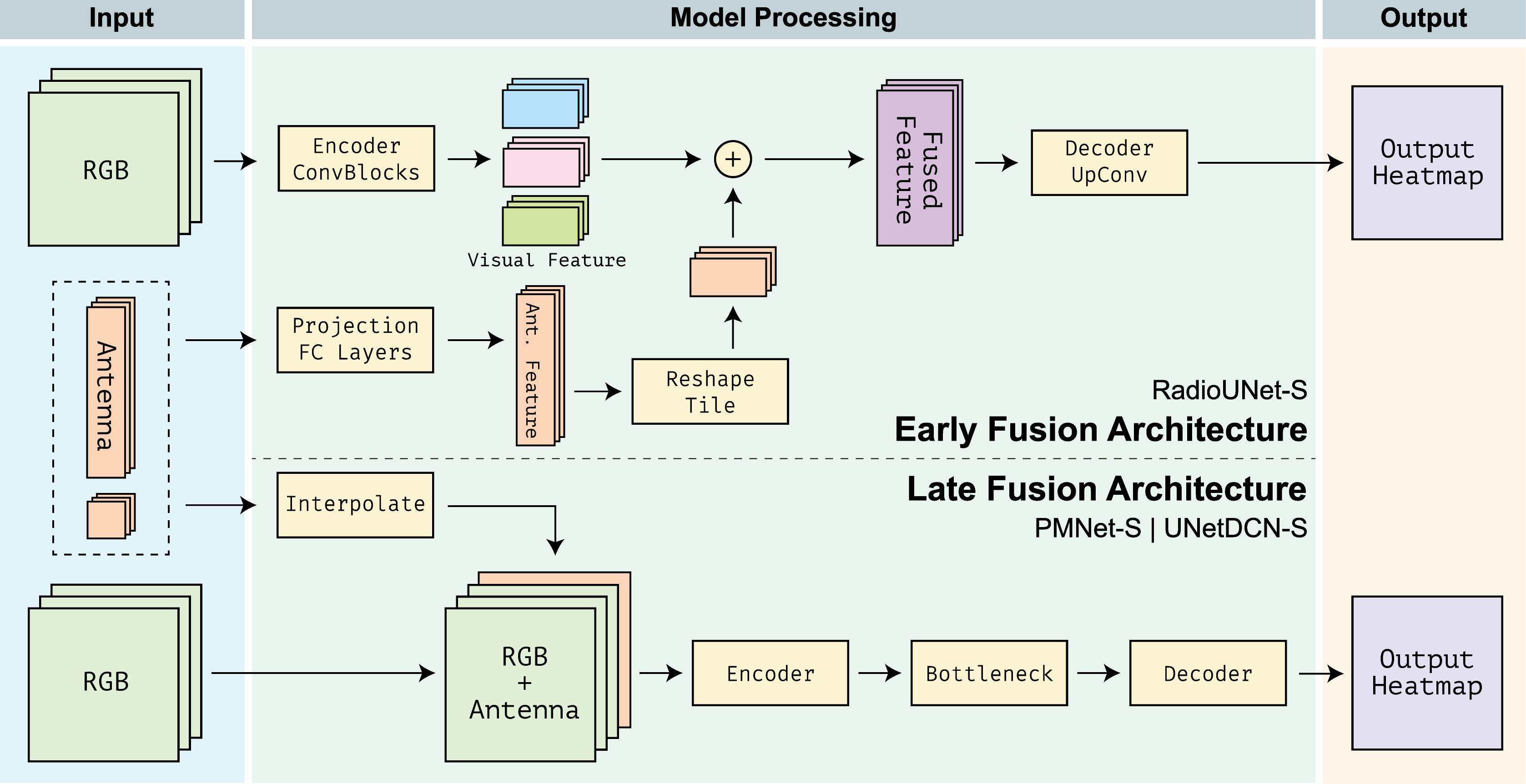}
    \caption{Modified architectures for existing radio map prediction methods adapted to the perspective-aligned radio mapping task.}
    \label{fig:appendix_architectures}
\end{figure*}

\par
\begin{table*}[htbp]
\centering
\caption{Comparison of SynthRM with Existing Datasets.}
\resizebox{\textwidth}{!}{%
\begin{tabular}{@{}lllllll c@{}}
    \toprule
    \textbf{Category} & \textbf{Dataset} & \textbf{Type} & \textbf{Visual Modality} & \textbf{Radio Label} & \textbf{Alignment} & \textbf{\makecell{Surface\\Aligned?}} & \textbf{Flexibility} \\ \midrule
    
    \makecell[l]{\textbf{I. Channel-Centric}\\(No Vision)} 
    & DeepMIMO \cite{deepmimo2019} & Synth. & \textit{None} & CSI Matrices & N/A & \xmark & High \\ \midrule
    
    \multirow{3}{*}{\makecell[l]{\textbf{II. Map-Based}\\(Pseudo-Visual)}} 
    & \makecell[l]{RadioMapSeer /\\2DRadioMap \cite{radiomapseer2022,radiomapseer3d2022}} & Synth. & Building Footprints & Top-down PL/ToA & Orthographic & \xmark & Low \\
    & CKMImageNet \cite{ckmimagenet2024} & Synth. & Building Footprints & Top-down CKM & Orthographic & \xmark & High \\
    & PMNet \cite{pmnet2024} & Synth. & \makecell[l]{2.5D Morphological\\Maps} & Top-down PL & Orthographic & \xmark & Low \\
    & RMDirectionalBerlin \cite{rmberlin2025} & Synth. & Aerial Orthophotos & Top-down PL & Orthographic & \xmark & Medium \\ \midrule
    
    \multirow{2}{*}{\makecell[l]{\textbf{III. Visual-Enabled}\\(Ill-Posed)}} 
    & ViWi \cite{viwi2020} & Synth. & \makecell[l]{Co-located RGB} & \makecell[l]{Top-down Beam/\\Power Map} & Positional Only & \xmark & Medium \\
    & DeepSense 5G \cite{deepsense6g2023} & Real & \makecell[l]{Multi-modal\\(RGB, LiDAR, Radar)} & \makecell[l]{Sparse Beam Index/\\Heatmap} & Positional Only & \xmark & Medium \\
    & M$^2$SC \cite{m3sc2023} & Synth. & \makecell[l]{Multi-modal\\(RGB-D, LiDAR, Radar)} & Waveform + CIR & Positional Only & \xmark & High \\ \midrule
    
    \textbf{\makecell[l]{IV. Proposed\\(Well-Posed)}} 
    & \textbf{SynthRM (Ours)} & \textbf{Synth.} & \textbf{\makecell[l]{RGB-D\\(Ego-centric)}} & \textbf{\makecell[l]{Perspective-Aligned\\Heatmap}} & \textbf{\makecell[l]{Pixel-Perfect\\Visible-Surface}} & \cmark & \textbf{High} \\ \bottomrule
\end{tabular}%
}
\label{tab:comparison}
\end{table*}

\subsection*{Model Modifications}
\label{sec:appendix_modifications}

In order to adapt existing radio map prediction methods to the perspective-aligned radio mapping task, we made several key modifications to the original architectures. The primary challenge lies in the input domain shift: original implementations of RadioUNet, PMNet, and UNetDCN were designed as image-to-image translation models where all inputs—such as city maps, transmitter masks, and terrain height maps—are presented as spatially aligned 2D tensors. In the context of SynthRM, however, critical signal determinants such as antenna patterns and transmitter configurations are provided as non-spatial feature vectors. To bridge this gap, we implement distinct fusion strategies tailored to the architectural characteristics of each baseline.

\subsubsection*{RadioUNet-S: Late Fusion via Bottleneck Injection}
For the RadioUNet-S adaptation, we modify the network topology to support a Late Fusion strategy. The standard RadioUNet architecture typically encodes transmitter information as a sparse spatial channel within the input tensor. In our implementation, we introduce a secondary input branch denoted as $x_{ant}$ to handle the dense parametric data of the transmitter. Rather than forcing this vector into the pixel space of the visual input, we employ a dedicated dense layer, $f_{proj}(\cdot)$, to project the antenna feature vector into the high-dimensional latent space of the network. This projected embedding is then fused directly into the U-Net's bottleneck via element-wise summation. This modification alters the internal information flow, allowing the network to modulate the high-level semantic features extracted from the visual encoder with the physical attributes of the transmitter before the decoding phase ensues.

\subsubsection*{PMNet-S and UNetDCN-S: Early Fusion via Spatial Broadcasting}
Conversely, for the PMNet-S and UNetDCN-S baselines, we adopt an Early Fusion strategy that preserves the original internal network topology. Instead of introducing new architectural branches, we modify the input representation using a spatial broadcasting technique. We utilize helper functions to spatially replicate the 1D antenna feature vector across the height and width dimensions, effectively inflating the vector into a 2D pseudo-image of matching spatial resolution. This broadcasted feature map is then concatenated channel-wise with the primary visual input (RGB-D) before entering the network. By instantiating the models with an adjusted input channel parameter, we treat the physical transmitter attributes as additional image channels. This allows the initial convolutional layers to implicitly learn the correlation between the global transmitter settings and local visual geometry without requiring structural changes to the encoder-decoder backbone.

The detailed architectures illustrating these fusion points are shown in Figure~\ref{fig:appendix_architectures}.

\section{Extended Analysis}
\label{sec:appendix_comparison}

Table~\ref{tab:comparison} presents a granular comparison of SynthRM against existing open-source resources, categorizing them by their geometric fidelity and alignment strategies. The comparison highlights three critical deficits in the current data landscape that SynthRM addresses:

\nosubsection{Category II - Geometric Causality}

Map-based datasets such as RadioMapSeer and PMNet rely on \textit{Allocentric} projections. While these datasets are sufficient for macroscopic coverage planning, the orthographic projection flattens vertical features, severing the causal link between visual occlusion and signal diffraction. As discussed in Section 2, this forces models to infer 3D signal behavior from 2D footprints, an inherently ill-posed task.

\nosubsection{Category III - Modal Cohesion}

Visual-enabled datasets like ViWi and DeepSense 6G introduce ego-centric imagery but often suffer from "Loose Coupling." In these datasets, the radio data is typically a heatmap or sparse beam index overlaid on a trajectory based solely on position. They lack \textit{Surface Alignment}, meaning the radio value at a specific pixel does not necessarily correspond to the object visible at that pixel, but rather to the volume of space the camera occupies.

\nosubsection{Category IV - Generative Scalability}

Unlike static recordings which are bound to fixed trajectories, or fixed-mesh simulations limited to a single topology, SynthRM functions as a procedural platform. This affords "High" flexibility, enabling the generation of infinite topological variations—a capability essential for training generalizable foundation models for 6G sensing.

\end{document}